\documentclass[twocolumn,showpacs,preprintnumbers,amsmath,amssymb]{revtex4}

\usepackage{graphicx}% Include figure files
\usepackage{dcolumn}% Align table columns on decimal point
\usepackage{bm}% bold math

\usepackage{color}

\usepackage{amssymb}

\usepackage{epsfig,enumerate}

\newcommand{\qH}{{\bf H}}
\newcommand{\qV}{{\bf\cal V}}
\newcommand{\qS}{{\bf S}}
\newcommand{\qw}{{\omega}}
\newcommand{\qwr}{{\hat\omega}}

\newcommand{\spec}{{\rm spec}}

\newcommand{\C}                 {{\bf C}}

\newcommand{\sY}{{\sigma(\bY)}}

\newcommand{\pa}        {\parallel}

\newcommand{\la}{{\langle}}
\newcommand{\ra}{{\rangle}}

\newcommand{\cW}{{\cal W}}

\newcommand{\cE}{{\cal E}}

\newcommand{\cN}{{\cal N}}
\newcommand{\cK}{{\cal K}}

\newcommand{\bP}{{\bf P}}
\newcommand{\bQ}{{\bf Q}}

\newcommand{\bu}{{\bf u}}

\newcommand{\bU}{{\bf U}}
\newcommand{\bY}{{\bf Y}}
\newcommand{\cL}{{\cal L}}
\newcommand{\cM}{{\cal M}}

\newcommand{\ds}{\displaystyle}

\newcommand{\be}{\begin{equation}}
\newcommand{\ee}{\end{equation}}
\newcommand{\ba}{\begin{array}}
\newcommand{\ea}{\end{array}}

\newcommand{\tr}{{\rm tr}}
\newcommand{\er}[1]{\hbox{(\ref{#1})}}

\newtheorem{theorem}            {Theorem}[section]

\newtheorem{lemma}              [theorem]{Lemma}
\newtheorem{sideremark}         [theorem]{Remark}
\newtheorem{sideeg}           [theorem]{Example}
\newtheorem{sideconj}           [theorem]{Conjecture}
\newtheorem{sideassumption}   [theorem]{Assumption}

\newenvironment{remark}         {\begin{sideremark}\rm}{\end{sideremark}}
\newenvironment{example}         {\begin{sideeg}\rm}{\end{sideeg}}

\def\argmin                     {\mathop{\rm argmin}}

\newcommand{\qed}               {\hfill $\square$}

\newcommand{\E}                 {{\bf E}}

\newcommand{\bydef}             {\stackrel{\triangle}{=}}
\newcommand{\inverse}[1]        {{\textstyle\frac{1}{#1}}}
\newcommand{\demi}              {\inverse{2}}

\date{November 3, 2003}

\begin{document}           % End of preamble and beginning of text.

%\preprint{APS/123-QED}

\title{Risk-Sensitive Optimal Control of Quantum Systems}

\author{M.R.~James}
\affiliation{Department of Engineering, Australian National University,
Canberra, ACT 0200,  Australia. \\
Matthew.James@anu.edu.au \\  quant-ph/0302136 }

%%%%%%%\maketitle                 % Produces the title.

\begin{abstract}
The importance of feedback control is being increasingly
appreciated in quantum physics and applications. This paper
describes the use of optimal control methods in the design of
quantum feedback control systems, and in particular the paper
formulates and solves a risk-sensitive optimal control problem.
The resulting risk-sensitive optimal control is given in terms of
a new unnormalized conditional state, whose dynamics include the
cost function used to specify the performance objective. The
risk-sensitive conditional dynamic equation describes the
evolution of our {\em knowledge} of the quantum system tempered by
our {\em purpose} for the controlled quantum system. Robustness
properties of risk-sensitive controllers are discussed, and an
example is provided.
\end{abstract}


\maketitle

\section{Introduction}
\label{sec:intro}

\begin{figure}[htb]
\begin{center}
\setlength{\unitlength}{1800sp}%
\begingroup\makeatletter\ifx\SetFigFont\undefined%
\gdef\SetFigFont#1#2#3#4#5{%
  \reset@font\fontsize{#1}{#2pt}%
  \fontfamily{#3}\fontseries{#4}\fontshape{#5}%
  \selectfont}%
\fi\endgroup%
\begin{picture}(7224,4749)(889,-4798)
\thinlines
{\color[rgb]{0,0,0}\put(3001,-1561){\framebox(3000,1500){}}
}%
{\color[rgb]{0,0,0}\put(2101,-3961){\framebox(2100,1500){}}
}%
{\color[rgb]{0,0,0}\put(4801,-3961){\framebox(2100,1500){}}
}%
{\color[rgb]{0,0,0}\put(4801,-3211){\vector(-1, 0){600}}
}%
{\color[rgb]{0,0,0}\put(2101,-3211){\line(-1, 0){1200}}
\put(901,-3211){\line( 0, 1){2400}}
\put(901,-811){\vector( 1, 0){2100}}
}%
{\color[rgb]{0,0,0}\put(6001,-811){\line( 1, 0){2100}}
\put(8101,-811){\line( 0,-1){2400}}
\put(8101,-3211){\vector(-1, 0){1200}}
}%
{\color[rgb]{0,0,0}\put(1801,-4786){\dashbox{60}(5400,2700){}}
}%
\put(5626,-3286){\makebox(0,0)[lb]{\smash{\SetFigFont{10}{14.4}{\familydefault}{\mddefault}{\updefault}{\color[rgb]{0,0,0}filter}%
}}}
\put(3301,-886){\makebox(0,0)[lb]{\smash{\SetFigFont{10}{14.4}{\familydefault}{\mddefault}{\updefault}{\color[rgb]{0,0,0}physical system}%
}}}
\put(1951,-1111){\makebox(0,0)[lb]{\smash{\SetFigFont{10}{14.4}{\familydefault}{\mddefault}{\updefault}{\color[rgb]{0,0,0}$u$}%
}}}
\put(6826,-1111){\makebox(0,0)[lb]{\smash{\SetFigFont{10}{14.4}{\familydefault}{\mddefault}{\updefault}{\color[rgb]{0,0,0}$y$}%
}}}
\put(2651,-3286){\makebox(0,0)[lb]{\smash{\SetFigFont{10}{14.4}{\familydefault}{\mddefault}{\updefault}{\color[rgb]{0,0,0}control}%
}}}
\put(3276,-4486){\makebox(0,0)[lb]{\smash{\SetFigFont{10}{14.4}{\familydefault}{\mddefault}{\updefault}{\color[rgb]{0,0,0}feedback controller}%
}}}
\put(1676,-661){\makebox(0,0)[lb]{\smash{\SetFigFont{10}{14.4}{\familydefault}{\mddefault}{\updefault}{\color[rgb]{0,0,0}input}%
}}}
\put(6551,-586){\makebox(0,0)[lb]{\smash{\SetFigFont{10}{14.4}{\familydefault}{\mddefault}{\updefault}{\color[rgb]{0,0,0}output}%
}}}
\end{picture}
\caption{Feedback controller showing the separation structure.}
\label{fig:fb}
\end{center}
\end{figure}
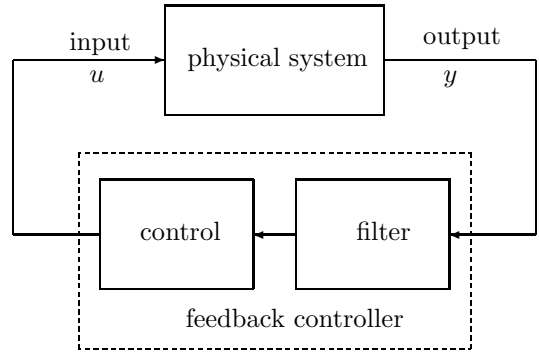

Optimal control theory provides a systematic approach to control
system design that is  widely used. A cost function is formulated
by the designer that encodes the desired performance of the system
as its minimum, and then the cost is minimized to obtain the
desired controller. Perhaps the most famous example is Kalman's
{\em linear quadratic Gaussian (LQG)} regulator problem, where the
cost criterion is an average of an integral,
\be
J^{LQG} = \E[\sum_{k=0}^{M-1} (x_k^\prime P x_k + u_k^\prime Q
u_k+x_M' P_M x_M)]
\label{lqg}
\ee
where $x_k$ and $u_k$ are respectively state and control variables
(vectors), and $P$, $Q$ and $P_M$ are weighting matrices
\footnote{If $x$ is a column vector, the notation $x'$ indicates the transpose, a row vector.}.
The cost criterion \er{lqg} is an example of what is sometimes
called a {\em risk-neutral} criterion. The state (or phase space)
variable $x_k$ is part of the model of the classical physical
system being controlled. In general, the controller has only
partial access to state information, with measurements corrupted
by noise.
 Kalman's optimal LQG feedback
controller is an explicit function of the conditional state and
covariance. It is dynamic, since the conditional state and
covariance evolve in time via the Kalman Filter (see, e.g.,
\cite{KA70},
\cite{KV86}), and the Kalman filter does not involve the cost function in
any way; it gives the optimal mean square state estimate
independently of any control objective. Interestingly, the
function giving the optimal feedback control is the same as for an
analogous problem with full state information, viz. multiplication
by a gain matrix determined by solving a Riccati equation.
Kalman's optimal LQG controller is the paradigm example of the
so-called {\em separation structure}, where the controller is
decomposed into an estimation part (filtering) and a control part,
as illustrated in Figure
\ref{fig:fb}.

%%%%%%%%%%%%%%%%%%%%%%%%%%%%%%%%%%%%%%%%%%%%%%%%%%%%%%%%%%%%%%
%%%%%%%%%%%%%%%%%%%%%%%%%%%%%%%%%%%%%%%%%%%%%%%%%%%%%%%%%%%%%%

 Over the past 20 or so years,
another type of optimal control problem has generated considerable
interest, viz. {\em linear exponential quadratic Gaussian (LEQG)}
optimal control, or {\em risk-sensitive} optimal control,
\cite{J73,W81,BV85}.  In this average of exponential of integral problem,
the cost is of the form
\be
\ba{l}
J^{LEQG}=
\\
\E[\exp\left( \sum_{k=0}^{M-1} (x_k^\prime P x_k +
u_k^\prime Q u_k)+x_M^\prime P_M x_M\right)] .
\ea
\label{leqg}
\ee
In this case the optimal feedback control is an explicit function
of a dynamical quantity closely related to the conditional state
and covariance, but given by dynamics that include terms from the
cost function. It also has a separation structure, though in this
case the filter depends on the cost function used to specify the
performance objective, and is a modification of the Kalman Filter.
One of the major reasons for the interest in the risk-sensitive
problem is its close connections to {\em robust control} and
minimax games,
\cite{GD88,DGKF89,JBE94}.
Robust control concerns the desire to design controllers that are
robust with respect to uncertainty, such as model errors and
exogenous disturbances, \cite{GL95}. Robustness properties of
risk-sensitive controllers are described in \cite{DJP00}.

Risk-neutral, risk-sensitive and other stochastic control problems
have been considered for problems with a finite number of states,
see, e.g. \cite{KV86,BJ97,FGM94,CM95}. After an analysis of an
example of a machine replacement problem \cite{CM95}, the authors
concluded that for that problem the risk-neutral controller was
more {\em aggressive} than risk-sensitive and related minimax
controllers.

Suppose we wish to control a quantum physical system using real
time feedback via a non-quantum  feedback system (say using a
digital computer) in some optimal fashion. If one were to do this
using a standard cost criterion, say one analogous to Kalman's
(LQG) regulator problem (risk-neutral), then one would find that
the optimal control is a function of the conditional (selective)
state (a density operator), as is well known, see, e.g.
\cite{VPB83,DJ99,DHJMT00,VDM01}.
The conditional state is the solution of a stochastic master
equation that describes the evolution of our {\em knowledge} of
the system. This stochastic master equation is used in two ways:
(i) as the model of the quantum physical system, taking into
account the effect of the measurements, and (ii) as the dynamics
of the filter in the optimal controller, Figure \ref{fig:fb}.

 The purpose of this
paper is to consider the {\em risk-sensitive} optimal control of
quantum physical systems. The quantum systems are modelled by
stochastic master equations for the conditional state. The
risk-sensitive criterion is one of a class of multiplicative cost
functions. The optimal solution for this class of problems has a
separation structure, Figure
\ref{fig:fb}, where the filter describes the evolution of an {\em
unnormalized} conditional state via a modified stochastic master
equation  that contains the cost function used to specify the
performance objective. The optimal control is a function of this
unnormalized conditional state. It is important to note that, in
contrast to the risk-neutral case described above, the states and
dynamics for the quantum physical model and the filter are not the
same. Indeed, the  unnormalized conditional dynamic equation used
in the filter describes the evolution of our {\em knowledge} of
the quantum system tempered by our {\em purpose} for the
controlled quantum system. This type of extension of the
conditional dynamics appears to be new to quantum physics, and may
merit further investigation. We emphasize that the  unnormalized
conditional state is defined only in the context of the
risk-sensitive and multiplicative control objectives considered
here, where it is used in a specific feedback situation. Again, we
emphasize that (i) the model of the quantum physical system is the
standard stochastic master equation for the conditional state, and
(ii) the filter is described by a modified stochastic master
equation  for an unnormalized conditional state; this modified
equation  includes terms from the cost function.

This paper is organized as follows. In section \ref{sec:cqs} we
carefully describe the model we use for the controlled quantum
system. Then in section \ref{sec:rn} we summarize some relevant
results for a risk-neutral optimal control problem, and make some
comments on the feedback solution. Section
\ref{sec:rs} contains the formulation and dynamic programming
solution to the risk-sensitive and related multiplicative cost
optimal control problems, together with a brief discussion of
robustness. The ideas are illustrated by a simple example of a
two-state system with feedback. Further developments, applications
and examples will be given in subsequent papers.

\noindent {\bf Acknowledgement.} The author wishes to thank Professor I.R.~Petersen
for many valuable discussions.

\section{The Controlled Quantum System}
\label{sec:cqs}

\subsection{Controlled State Transfer}
\label{sec:cst}

 We consider a controlled quantum physical system with {\em inputs} $u$ and {\em outputs} $y$.
The inputs represent signals or actions that are applied to the
system, such as voltages, forces, or light pulses. The outputs are
signals that result from repeated measurements of observable
quantities,  such as position, spin, etc. We will assume, for
simplicity, that the measurements are {\em discrete} valued. It is
sometimes useful to denote the range of input and output values by
$\bU$ and $\bY$ respectively.

The {\em state} of the quantum system is described by a density
operator $\qw$\footnote{Without further qualification, we use the
term state to refer to a positive self-adjoint operator normalized
to have trace equal to one.}.
 This state evolves in time as a result of a
variety of factors including the underlying unitary evolution,
interaction with the environment, the effect of repeated
measurements, and feedback control actions. Since measurements are
made, and the outcomes are used to determine control actions in a
feedback context, we are interested in the {\em selective} or {\em
conditional} evolution of the states. As an example
\cite{BB91,VPB01,WM94}, a range of conditional evolutions can be described
by an Ito-type {\em stochastic master equations (SME)} of the form
\be
d \qw  = \cL [\qw] dt + \cM[\qw] dW
\label{sme}
\ee
for suitable (super) operators $\cL$ and
$\cM$ (which may depend on the control $u$). Here, $dW$ represents
an Ito-type Brownian motion (Wiener process) increment, called an
innovation, related to the measured output value $y$ by
\be
dy = \tr\{ \cN[\qw] \} dt + dW
\label{sme-y}
\ee
for a suitable (super) operator $\cN$. If we denote by $\rho$ the
expected value of
$\qw$ with respect to
$W$ (or $y$), we obtain the {\em master equation}, frequently
encountered in the analysis of open systems:
\be
\dot \rho  = \cL[\rho] .
\label{me}
\ee

It is conceptually and technically simpler to work in {\em
discrete time}, and so we will do so in this paper. Effectively,
we will be using a model for sampled-data feedback control of
quantum systems. In this model, measurements are made and control
actions are applied at discrete time instants $t_k$\footnote{For
example, $t_k = k \Delta t$, where $\Delta t$ is the time between
samples, and $k$ is an integer indicating discrete time values.},
 called sample
times. Continuous time models are of considerable importance, and
will be considered elsewhere.

The discrete time model we use for the quantum system is defined
in terms of a (super) operator $\Gamma(u,y)$
\footnote{The operator $\Gamma(u,y)$ should preserve self-adjointness and positivity.}
that depends on the control input $u$ and the output measurement
$y$. The idea is that if the quantum system is in state $\qw_k$ at
time $k$, and at this time the control value $u_k$ is applied, a
measurement outcome $y_{k+1}$ will be recorded, and the system
will transfer to a new state $\qw_{k+1}$.
 The
probability of $y_{k+1}$ is $p(y_{k+1} \vert u_k, \qw_k)$, where
\be
p(y \vert u, \qw) = \la \Gamma(u,y)\qw, I \ra .
\label{trans-prob}
\ee
Here, we have used the notation
\be
\la \qw, B \ra = \tr [ B \qw ]
\label{tr-ip}
\ee
to specify the (expected) value of an observable $B$ when the
system is in state $\qw$. The operator $\Gamma(u,y)$ is assumed to
be normalized, i.e.
$$
\sum_{y \in \bY}  \la \Gamma(u,y)\qw, I \ra =  \la \qw, I \ra =1
$$
so that $p(y \vert u, \qw)$ is a {\em probability distribution},
since it satisfies $\sum_{y}p(y \vert u, \qw) = 1$.

Selective or conditional evolution means that the new state
$\qw_{k+1}$ depends on the value of the measurement $y_{k+1}$, and
we write this dependance as follows:
\be
 \qw_{k+1} = \Lambda_\Gamma(u_k, y_{k+1})  \qw_k,
\label{system}
\ee
where
\be
\Lambda_\Gamma(u,y)\qw = \ds
\frac{\Gamma(u,y)\qw}{ p(y \vert u, \qw)} .
\label{gamma-dagger-bar}
\ee
Equation \er{system} is a discrete time {\em stochastic master
equation (SME)}, and can be viewed, e.g., as the result of
integrating an equation of the form \er{sme} over one time step
(after substituting for $dW$ in terms of $dy$).

 We denote the
average of the conditional state $\qw_k$ with respect to the
measurements by
$\rho_k$. If $u_k$ is a deterministic (non-random) input signal,
then $\rho_k$ satisfies the {\em master equation}
\be
\rho_{k+1} = \sum_{y \in \bY} \Gamma(u_k,y) \rho_k .
\label{system-me}
\ee

{\em Equation \er{system} constitutes our model of the quantum
system.} For further information on this framework of operator
valued measures and quantum operations, see
\cite{EBD76,NC00,GZ00,VPB83}. We now give some examples.

\begin{example} \label{eg:new} We define the controlled transfer $\Gamma(u,y)$ by interleaving
open system dynamics and imperfect orthogonal measurements. The
open system dynamics are modelled by a quantum operation
\be
\cE^u \qw = \sum_b E_b^u \qw  E_b^{u \, \dagger}
\label{open-1}
\ee
where the controlled operators $E_b^u$ satisfy $\sum_b E^{u
\,
\dagger}_b E^u_b
= I$ for all inputs $u$. Closed systems are described by the
unitary evolution operation $\cE^u
\qw = T^u \qw T^{u \, \dagger}$, where for each
input value $u$, $T^u$ is a unitary operator.

The imperfect measurements are modelled as follows. Let $A$ be a
self-adjoint operator with discrete nondegenerate spectrum
$\spec(A)$. For $a\in\spec(A)$ an eigenvalue of $A$ let $\vert a
\ra$ denote the normalized eigenvector, and let $P_a =
\vert a \ra \la a \vert$ denote the projection onto the eigenspace
of $A$ ($P_a \vert \psi \ra = \la a
\vert \psi \ra \vert a \ra$). Perfect measurements would
correspond to $y=a$; however, to reflect the presence of
measurement noise in applications we will assume that when a
measurement occurs on the quantum system, the values $a$ and
associated projections occur in the usual (perfect) way, but that
knowledge of the outcomes is corrupted by sensor noise so that the
controller (or any observing device or person) measures a value
$y$.  The measurement $y$ is a random variable, related to the
outcomes $a$ via probability kernels $q(y\vert a)$, the
probability of $y$ given that $a$ occurred. The kernels have the
property that $\sum_y q(y
\vert a) =1$ for all $a$. In the case of perfect measurements,
$q(y\vert a) =1$ if $y=a$, and $q(y\vert a) =0$ if $y\neq a$.

The operator $\Gamma(u,y)$ is given by
\be
\Gamma(u,y) \qw = \sum_{a,b} q(y \vert a)  P_a  E_b^u\qw E_b^{u \dagger} P_a
\label{gamma-dist-imperfect}
\ee
and the adjoint is given by
\be
\Gamma^\dagger(u,y) B = \sum_{a,b} q(y \vert a)   E_b^{u \dagger} P_a B P_aE_b^u
\label{gamma-dist-imperfect-dagger}
\ee
where $B$ is an observable. The expressions in this example can be
derived using standard techniques of quantum operations and
discrete time filtering based on Bayes' Rule (see, e.g.,
\cite[Chapter 2.2]{GZ00}, \cite[Chapter 8]{NC00},
\cite{EBD76}, \cite{VPB83},
\cite[Chapter 6]{KV86}, \cite[Chapter 7]{KA70}).
\qed
\end{example}

\begin{example} \label{eg:2-d-1}(Two-state system.)
We now describe a specific instance of Example
\ref{eg:new}, viz. a two-state system and measurement device, where it is desired to use feedback control to
put the system into a given state. The example is inspired by a
simple quantum feedback example
\cite[Section 1.3]{HW94} and an example in stochastic control concerning a machine
replacement problem \cite{FGM94,CM95}.

In \cite[Section 1.3]{HW94}, a particle beam is passed through a
Stern-Gerlach device, which results in one beam of particles in
the up state, and one beam in the down state. The beam of
particles in the up state is subsequently left alone, while the
beam in the down state is subject to a further device which will
result in a change of spin direction from down to up. The final
outcome of this feedback arrangement is that all particles are in
the up state. Analogous feedback configurations can be constructed
using other physical systems, e.g. light and polarization
measurement.

In what follows we extend the general  features of this example to
accommodate repeated noisy measurements. Physically, the noisy
measurements  might arise from imperfectly separated beams, where
a proportion of each beam contaminates the other, and/or from
interference or noise affecting sensors. The example was chosen
because the risk-neutral and risk-sensitive problems can be solved
explicitly. Hence the example provides a concrete illustration of
some ideas concerning quantum feedback control. More substantial
examples and applications will be considered elsewhere.

The pure states of the system are of the form
$$
\vert \psi \ra = c_{-1} \vert -1 \ra + c_1 \vert 1 \ra \ \equiv \
\left( \ba{c} c_{-1} \\ c_1 \ea \right) .
$$
The states $\vert -1 \ra$ and $\vert 1 \ra$ are eigenstates of the
observable
\be
A = \left( \ba{cc} -1 & 0 \\ 0 & 1
\ea
\right)
\label{obs-A}
\ee
corresponding to ideal measurement values $a=-1$ and $a=1$. It is
desired to put the system into the state
$$
\vert 1 \ra = \left( \ba{c} 0 \\ 1 \ea  \right),
\  \text{or}  \ \vert 1 \ra  \la 1 \vert = \left( \ba{cc} 0 & 0 \\ 0 & 1 \ea
\right).
$$

%%%%%%%%%%%%%%%%%%%%%%%%%%%%%%%%%%%%%%%%%%%%%%%%%%%%%%%%%%%%%%
\begin{figure}[htb]
\begin{center}
\setlength{\unitlength}{1800sp}%

\begingroup\makeatletter\ifx\SetFigFont\undefined%
\gdef\SetFigFont#1#2#3#4#5{%
  \reset@font\fontsize{#1}{#2pt}%
  \fontfamily{#3}\fontseries{#4}\fontshape{#5}%
  \selectfont}%
\fi\endgroup%
\begin{picture}(7224,2424)(889,-3073)
\thinlines
{\color[rgb]{0,0,0}\put(2101,-2161){\framebox(1800,1500){}}
}%
{\color[rgb]{0,0,0}\put(5101,-2161){\framebox(1800,1500){}}
}%
{\color[rgb]{0,0,0}\put(901,-1411){\vector( 1, 0){1200}}
}%
{\color[rgb]{0,0,0}\put(3901,-1411){\vector( 1, 0){1200}}
}%
{\color[rgb]{0,0,0}\put(3001,-3061){\vector( 0, 1){900}}
}%
{\color[rgb]{0,0,0}\put(6001,-2161){\vector( 0,-1){900}}
}%
{\color[rgb]{0,0,0}\put(6901,-1411){\vector( 1, 0){1200}}
}%
\put(1126,-1786){\makebox(0,0)[lb]{\smash{\SetFigFont{10}{14.4}{\familydefault}{\mddefault}{\updefault}{\color[rgb]{0,0,0}$\qw_k$}%
}}}
\put(2926,-1561){\makebox(0,0)[lb]{\smash{\SetFigFont{10}{14.4}{\familydefault}{\mddefault}{\updefault}{\color[rgb]{0,0,0}$T^u$}%
}}}
\put(3151,-2836){\makebox(0,0)[lb]{\smash{\SetFigFont{10}{14.4}{\familydefault}{\mddefault}{\updefault}{\color[rgb]{0,0,0}$u_k$}%
}}}
\put(5701,-1486){\makebox(0,0)[lb]{\smash{\SetFigFont{10}{14.4}{\familydefault}{\mddefault}{\updefault}{\color[rgb]{0,0,0}M-$\alpha$}%
}}}
\put(6151,-2836){\makebox(0,0)[lb]{\smash{\SetFigFont{10}{14.4}{\familydefault}{\mddefault}{\updefault}{\color[rgb]{0,0,0}$y_{k+1}$}%
}}}
\put(7201,-1786){\makebox(0,0)[lb]{\smash{\SetFigFont{10}{14.4}{\familydefault}{\mddefault}{\updefault}{\color[rgb]{0,0,0}$\qw_{k+1}$}%
}}}
\end{picture}

\caption{Two-state system  example showing the controlled unitary operator $T^u$ and the noisy measurement device M-$\alpha$ with error probability $\alpha$.}
\label{fig:sg}
\end{center}
\end{figure}
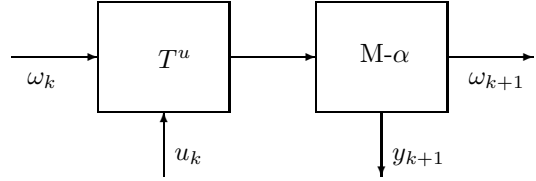
%%%%%%%%%%%%%%%%%%%%%%%%%%%%%%%%%%%%%%%%%%%%%%%%%%%%%%%%%%%%%%

We define a controlled transfer operator $\Gamma(u,y)$ as the
following physical process, Figure \ref{fig:sg}. First apply a
unitary transformation $T^u$, where the control value $u=0$ means
do nothing, while $u=1$ means to flip the states (quantum not
gate), i.e.
$$
T^u = \left\{ \ba{rl}
\left( \ba{cc} 1 & 0 \\ 0 & 1 \ea \right) & \ \text{if} \ u=0
\\
\left( \ba{cc} 0 & 1 \\ 1 & 0 \ea \right) & \ \text{if} \ u=1 .
\ea \right.
$$
We then make an imperfect  measurement corresponding to the
observable $A$. We model this by an ideal device (e.g.
Stern-Gerlach, beam splitter) with projection operators
$$
P_{-1} = \left( \ba{cc} 1 & 0 \\ 0 & 0 \ea \right), \ P_{1} =
\left( \ba{cc} 0 & 0 \\ 0 & 1 \ea \right)
$$
followed by a memoryless channel with error probability kernels
$$
\ba{ll}
q(-1 \vert -1 ) & = 1 - \alpha
\\
q(-1 \vert 1 ) & =  \alpha
\\
q(1 \vert -1 ) & =  \alpha
\\
q(1 \vert 1 ) & = 1 - \alpha
\ea
$$
where $0\leq \alpha \leq 1$ is the probability of a measurement
error  (cf. \cite[Figure 8.1]{NC00}).

The controlled transfer operator is therefore (from
\er{gamma-dist-imperfect})
$$
\ba{l}
\Gamma(u,y)\qw = q(y\vert -1) P_{-1} T^u \qw T^{u\,\dagger} P_{-1} \\
\hspace{2.0cm} +
q(y\vert 1) P_{1} T^u \qw T^{u\,\dagger} P_{1} .
\ea
$$
In this example, the control $u$ can take the values $0$ or $1$,
and output $y$ has values $0$ or $1$ ($\bU=\{ 0,1\})$, $\bY=\{
0,1\})$.

If we write a general density matrix as
\be
\qw = \left( \ba{cc} \qw_{11} & \qw_{12} \\ \qw_{12}^\ast & \qw_{22} \ea
\right),
\label{density-2d}
\ee
then the controlled operators $\Gamma(u,y)$ are given explicitly
by
$$
\ba{rl}
\Gamma(0,-1)\qw & = \left( \ba{cc} (1-\alpha)\qw_{11} & 0 \\ 0 &  \alpha \qw_{22} \ea \right)
\\
\Gamma(0,1)\qw & = \left( \ba{cc} \alpha \qw_{11} & 0 \\ 0 &  (1-\alpha) \qw_{22} \ea \right)
\\
\Gamma(1,-1)\qw & = \left( \ba{cc} (1-\alpha)\qw_{22} & 0 \\ 0 &  \alpha \qw_{11} \ea \right)
\\
\Gamma(1,1)\qw & = \left( \ba{cc} \alpha\qw_{22} & 0 \\ 0 &  (1-\alpha) \qw_{11} \ea \right)
\ea
$$
This example is  continued in stages in the remainder of the paper
(Examples \ref{eg:2-d-2}, \ref{eg:2-d-3}, \ref{eg:2-d-4},
\ref{eg:2-d-5}).
\qed
\end{example}

\subsection{Feedback Control}
\label{sec:cd}

In the above description of the quantum system \er{system}, we
have not described how the controls $u_k$ are determined by  the
measurements $y_k$ via a feedback controller $K$. We now do this.

Feedback controllers should be {\em causal}, i.e., the current
control value $u_k$ cannot depend on future values of the
measurements $y_{k+1}, y_{k+2}, \ldots$. On a time interval $0
\leq k \leq M-1$ this is expressed as follows:
$$
K = \{ K_{0}, K_{1}, \ldots, K_{M-1} \}
$$
where
$$
\ba{rl}
u_0 & = K_0
\\
u_1 & = K_1(y_1)
\\
u_2 & = K_2(y_1,y_2)
\\
\text{etc.}
\ea
$$

%%%%%%%%%%%%%%%%%%%%%%%%%%%%%%%%%%%%%%%%%%%%%%%%%%%%%%%%%%%%%%
\begin{figure}[htb]
\begin{center}
\setlength{\unitlength}{1800sp}%
\begingroup\makeatletter\ifx\SetFigFont\undefined%
\gdef\SetFigFont#1#2#3#4#5{%
  \reset@font\fontsize{#1}{#2pt}%
  \fontfamily{#3}\fontseries{#4}\fontshape{#5}%
  \selectfont}%
\fi\endgroup%
\begin{picture}(7224,3924)(889,-3973)
\thinlines
{\color[rgb]{0,0,0}\put(3001,-1561){\framebox(3000,1500){}}
}%
{\color[rgb]{0,0,0}\put(2101,-3211){\line(-1, 0){1200}}
\put(901,-3211){\line( 0, 1){2400}}
\put(901,-811){\vector( 1, 0){2100}}
}%
{\color[rgb]{0,0,0}\put(6001,-811){\line( 1, 0){2100}}
\put(8101,-811){\line( 0,-1){2400}}
\put(8101,-3211){\line(-1, 0){1200}}
}%
{\color[rgb]{0,0,0}\put(3451,-3961){\framebox(2100,1500){}}
}%
{\color[rgb]{0,0,0}\put(6901,-3211){\vector(-1, 0){1350}}
}%
{\color[rgb]{0,0,0}\put(3451,-3211){\line(-1, 0){1350}}
}%
\put(3301,-886){\makebox(0,0)[lb]{\smash{\SetFigFont{10}{14.4}{\familydefault}{\mddefault}{\updefault}{\color[rgb]{0,0,0}physical system}%
}}}
\put(1951,-1111){\makebox(0,0)[lb]{\smash{\SetFigFont{10}{14.4}{\familydefault}{\mddefault}{\updefault}{\color[rgb]{0,0,0}$u$}%
}}}
\put(6826,-1111){\makebox(0,0)[lb]{\smash{\SetFigFont{10}{14.4}{\familydefault}{\mddefault}{\updefault}{\color[rgb]{0,0,0}$y$}%
}}}
\put(1676,-661){\makebox(0,0)[lb]{\smash{\SetFigFont{10}{14.4}{\familydefault}{\mddefault}{\updefault}{\color[rgb]{0,0,0}input}%
}}}
\put(6551,-586){\makebox(0,0)[lb]{\smash{\SetFigFont{10}{14.4}{\familydefault}{\mddefault}{\updefault}{\color[rgb]{0,0,0}output}%
}}}
%\put(3751,-3061){\makebox(0,0)[lb]{\smash{\SetFigFont{10}{14.4}{\familydefault}{\mddefault}{\updefault}{\color[rgb]{0,0,0}feedback controller}%
%}}}
\put(4351,-3436){\makebox(0,0)[lb]{\smash{\SetFigFont{10}{14.4}{\familydefault}{\mddefault}{\updefault}{\color[rgb]{0,0,0}$K$}%
}}}
\end{picture}

\caption{Feedback control of quantum system showing a general feedback controller $K$.}
\label{fig:fb-q}
\end{center}
\end{figure}
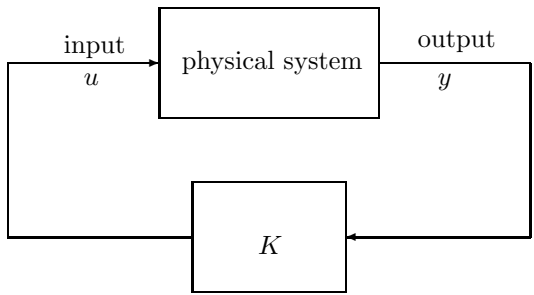
%%%%%%%%%%%%%%%%%%%%%%%%%%%%%%%%%%%%%%%%%%%%%%%%%%%%%%%%%%%%%%

To simplify notation, we often write sequences
$u_{k_1},u_{k_1+1},\ldots,u_{k_2}$ as $u_{k_1,k_2}$. Then we can
write $u_k = K_k(y_{1,k})$. A controller $K$ can be restricted to
subintervals $k
\leq j \leq M$ by fixing (or omitting) the first arguments in the
obvious way. We denote by $\cK$ the class of all such feedback
controllers.

A feedback controller $K$ in closed loop with the quantum system,
Figure \ref{fig:fb-q},  operates as follows. The given  initial
state $\qw_{0}$ and  controller $K$ are sufficient to define
random sequences of states $\qw_{0,M}$, inputs $u_{0,M-1}$ and
outputs $y_{1,M}$ over a given time interval $0 \leq k \leq M$
iteratively as follows. The control value $u_{0}$ is determined by
$K_0$   (no observations are involved yet), and it is applied to
the quantum system, which responds by selecting $y_{1}$ at random
according to the distribution $p(y_{1}
\vert u_{0},
\qw_{0})$. This then determines the next state $\qw_{1}$ via \er{system}. Next
$u_{1}$ is given by $K_1(y_1)$, and applied to the system. This
process is repeated until the final time.

The controller $K$ therefore determines controlled stochastic
processes $\qw_k$, $u_k$ and $y_k$ on the interval $0 \leq k \leq
M$. Expectation with respect to the associated probability
distribution is denoted $\E_{\qw_0,0}^K$. The state sequence
$\qw_k$ is a {\em controlled Markov process}.

One way a controller $K$ can be constructed is using a function
$$
u_k = \bu(\qw_k,k)
$$
where $\qw_k$ is given by \er{system} with initial state $\qw_0$.
This controller is denoted $K^{\bu}_{\qw_0}$. The SME equation
\er{system} forms part of this controller, viz. its dynamics, and
must be implemented with suitable technology (e.g. digital
computer).  Controllers of this type are said to have a {\em
separation structure}, where the controller can be decomposed into
an estimation part (i.e. filtering via
\er{system}) and a control part (i.e. the function $\bu$). We will
see in section \ref{sec:rn} that the optimal risk-neutral
controller is of this form (Figure
\ref{fig:fb-rn}). In section \ref{sec:rs}, the
optimal risk-sensitive controller also has a separation structure,
but the filter used is different (Figure
\ref{fig:fb-rs}). The separation structure arises
naturally from the dynamic programming techniques, as we shall
see.

\begin{example} \label{eg:2-d-2} (Two-state system with feedback, Example \ref{eg:2-d-1} continued.)
We consider a particular feedback controller $\bar K$ for a time
horizon
$M=2$ defined by
\be
u_0=\bar K_0 = 0, \  \ \  u_1=\bar K_1(y_1) = \left\{  \ba{rl} 0 &
\
\text{if} \ y_1=1
\\
1 & \
\text{if} \ y_1=-1 .
\ea
\right.
\label{eg1-bar-K}
\ee
We apply $\bar K$ to the system with initial pure state
\be
\vert \psi_0 \ra = \frac{1}{\sqrt{2}} \vert -1 \ra +  \frac{1}{\sqrt{2}} \vert 1 \ra
, \ \text{or} \ \qw_0 = \demi \left( \ba{cc} 1 & 1 \\ 1 & 1 \ea
\right) .
\label{eg1-w0}
\ee

\begin{table}
$$
\begin{array}{|c|cc|cc|}
\hline
\qw_0 & p_1 & \qw_1 & p_2 &  \qw_2  \\
\hline
& & & 2\alpha(1-\alpha) & \qw_2^{(0,-1),(1,-1)} \\
\qw_0 & \demi &\qw_1^{(0,-1)}  & \alpha^2+(1-\alpha)^2 & \qw_2^{(0,-1),(1,1)} \\
& \demi & \qw_1^{(0,1)}  & 2\alpha(1-\alpha) & \qw_2^{(0,1),(0,-1)}  \\
& & & \alpha^2+(1-\alpha)^2 & \qw_2^{(0,1),(0,1)}  \\
\hline
\rho_0=\qw_0 &   & \rho_1 &  & \rho_2 \\
\hline
\end{array}
$$
\label{table:eg-1}
\caption{State evolution under the controller $\bar K$.}
\end{table}

The result is shown in Table \ref{table:eg-1}, which displays the
resulting conditional states
$$
\ba{rl}
\qw_1^{(u_0,y_1)} & = \Lambda_\Gamma(u_0,y_1)\qw_0,
\\
\qw_2^{(u_0,y_1),(u_1,y_2)} & = \Lambda_\Gamma(u_1,y_2)\qw_1^{(u_0,y_1)}
\ea
$$
and the associated probabilities.
 Explicitly, the terms shown in Table \ref{table:eg-1} are:
$$
p_1 = p(y_1 \vert u_0, \qw_0), \  \  p_2 = p(y_2 \vert u_1, \qw_1)
$$
$$
\qw_1^{(0,-1)} =  \left( \ba{cc} (1-\alpha) & 0 \\ 0 & \alpha \ea
\right), \  \
\qw_1^{(0,1)} = \left( \ba{cc} \alpha & 0 \\ 0 & (1-\alpha) \ea
\right)
$$
$$
\qw_2^{(0,-1),(1,-1)} = \qw_2^{(0,1),(0,-1)} = \demi \left( \ba{cc} 1 & 0 \\ 0 & 1 \ea
\right),
$$
$$
\ba{l}
\qw_2^{(0,-1),(1,1)} = \qw_2^{(0,1),(0,1)}
\\ \hspace{2.0cm}  = \ds{\frac{1}{\alpha^2 + (1-\alpha)^2}\left( \ba{cc} \alpha^2 & 0 \\ 0 & (1-\alpha)^2 \ea
\right)} .
\ea
$$
Also shown are the non-selective states:
$$
\rho_0 = \qw_0
$$
$$
\ba{rl}
\rho_1 & = p(-1 \vert u_0, \qw_0) \qw_1^{(0,-1)} + p(1 \vert u_0, \qw_0) \qw_1^{(0,1)}
\\
& = \ds{\demi \left( \ba{cc} 1 & 0 \\ 0 & 1 \ea \right)}
\ea
$$
\be
\ba{l}
\rho_2  =
 p(-1 \vert 1, \qw_1^{(0,-1)}) \qw_2^{(0,-1),(1,-1)}
\\ \hspace{1.0cm} + p(1 \vert 1, \qw_1^{(0,-1)}) \qw_2^{(0,-1),(1,1)}
\\
  \hspace{1.0cm}  +  p(-1 \vert 0, \qw_1^{(0,1)}) \qw_2^{(0,1),(0,-1)}
 \\
 \hspace{1.0cm} + p(1
\vert 0, \qw_1^{(0,1)}) \qw_2^{(0,1),(0,1)}
\\
\
= \demi \left( \ba{cc} \alpha^2 +\alpha(1-\alpha) & 0 \\ 0 &
\alpha(1-\alpha)+(1-\alpha)^2 \ea \right) .
\ea
\label{eg1-rho-2}
\ee

At time $k=0$ the control $u=0$ is applied. If $y_1=-1$ is
observed, as a result of the imperfect measurement, the system
moves to the state $\qw_1^{(0,-1)}$. Since $y_1=-1$, the
controller $\bar K$ \er{eg1-bar-K} gives $u_1=1$. This results in
the states $\qw_2^{(0,-1),(1,-1)}$ or  $\qw_2^{(0,-1),(1,1)}$,
depending on the outcome of the second measurement $y_2$. If, on
the other hand,  $y_1=1$ is observed, the system moves to the
state $\qw_1^{(0,1)}$. Since $y_1=1$, the controller $\bar K$
\er{eg1-bar-K} gives $u_1=0$, and hence
$\qw_2^{(0,1),(0,-1)}$ or  $\qw_2^{(0,1),(0,1)}$, again depending
on the outcome of the second measurement $y_2$. This is
illustrated in Figure \ref{fig:sg-2}.

%%%%%%%%%%%%%%%%%%%%%%%%%%%%%%%%%%%%%%%%%%%%%%%%%%%%%%%%%%%%%%
\begin{figure}[htb]
\begin{center}
\setlength{\unitlength}{1600sp}%
\begingroup\makeatletter\ifx\SetFigFont\undefined%
\gdef\SetFigFont#1#2#3#4#5{%
  \reset@font\fontsize{#1}{#2pt}%
  \fontfamily{#3}\fontseries{#4}\fontshape{#5}%
  \selectfont}%
\fi\endgroup%
\begin{picture}(8724,2470)(2389,-4019)
\thinlines
{\color[rgb]{0,0,0}\put(8401,-2761){\framebox(900,900){}}
}%
{\color[rgb]{0,0,0}\put(6526,-2986){\framebox(750,675){}}
}%
{\color[rgb]{0,0,0}\put(6001,-3361){\dashbox{60}(4050,1800){}}
}%
{\color[rgb]{0,0,0}\put(3751,-2761){\framebox(900,900){}}
}%
{\color[rgb]{0,0,0}\put(4651,-2611){\vector( 1, 0){1875}}
}%
{\color[rgb]{0,0,0}}%
{\color[rgb]{0,0,0}}%
{\color[rgb]{0,0,0}\put(7276,-2611){\line( 1, 0){525}}
\put(7801,-2611){\vector( 0, 1){300}}
}%
{\color[rgb]{0,0,0}\put(4651,-2011){\line( 1, 0){3150}}
\put(7801,-2011){\vector( 0,-1){300}}
}%
{\color[rgb]{0,0,0}\put(7801,-2311){\vector( 1, 0){600}}
}%
{\color[rgb]{0,0,0}\put(2401,-2311){\vector( 1, 0){1350}}
}%
{\color[rgb]{0,0,0}}%
{\color[rgb]{0,0,0}\put(3001,-3361){\dashbox{60}(2400,1800){}}
}%
{\color[rgb]{0,0,0}\put(9301,-2611){\line( 1, 0){1350}}
\put(10651,-2611){\vector( 0, 1){300}}
}%
{\color[rgb]{0,0,0}\put(9301,-2011){\line( 1, 0){1350}}
\put(10651,-2011){\vector( 0,-1){300}}
}%
{\color[rgb]{0,0,0}\put(10651,-2311){\vector( 1, 0){450}}
}%
{\color[rgb]{0,0,0}}%
\put(8476,-2386){\makebox(0,0)[lb]{\smash{\SetFigFont{8}{14.4}{\familydefault}{\mddefault}{\updefault}{\color[rgb]{0,0,0}M-$\alpha$}%
}}}
\put(6651,-2686){\makebox(0,0)[lb]{\smash{\SetFigFont{10}{14.4}{\familydefault}{\mddefault}{\updefault}{\color[rgb]{0,0,0}flip}%
}}}
\put(3826,-2386){\makebox(0,0)[lb]{\smash{\SetFigFont{8}{14.4}{\familydefault}{\mddefault}{\updefault}{\color[rgb]{0,0,0}M-$\alpha$}%
}}}
\put(5476,-3961){\makebox(0,0)[lb]{\smash{\SetFigFont{10}{14.4}{\familydefault}{\mddefault}{\updefault}{\color[rgb]{0,0,0}$\qw_1$}%
}}}
\put(4876,-2436){\makebox(0,0)[lb]{\smash{\SetFigFont{10}{14.4}{\familydefault}{\mddefault}{\updefault}{\color[rgb]{0,0,0}$1$}%
}}}
\put(9526,-2436){\makebox(0,0)[lb]{\smash{\SetFigFont{10}{14.4}{\familydefault}{\mddefault}{\updefault}{\color[rgb]{0,0,0}$1$}%
}}}
\put(4876,-3036){\makebox(0,0)[lb]{\smash{\SetFigFont{10}{14.4}{\familydefault}{\mddefault}{\updefault}{\color[rgb]{0,0,0}$-1$}%
}}}
\put(9526,-3036){\makebox(0,0)[lb]{\smash{\SetFigFont{10}{14.4}{\familydefault}{\mddefault}{\updefault}{\color[rgb]{0,0,0}$-1$}%
}}}
\put(2401,-3961){\makebox(0,0)[lb]{\smash{\SetFigFont{10}{14.4}{\familydefault}{\mddefault}{\updefault}{\color[rgb]{0,0,0}$\qw_0$}%
}}}
\put(10576,-3961){\makebox(0,0)[lb]{\smash{\SetFigFont{10}{14.4}{\familydefault}{\mddefault}{\updefault}{\color[rgb]{0,0,0}$\rho_2$}%
}}}
\put(9901,-3961){\makebox(0,0)[lb]{\smash{\SetFigFont{10}{14.4}{\familydefault}{\mddefault}{\updefault}{\color[rgb]{0,0,0}$\tilde\qw_2$}%
}}}
\end{picture}

\caption{Physical realization of the two stages of the two-state system with feedback using controller $\bar K$.
Due to the merging of the beams in the second stage, we have the
intermediate state $\tilde\qw_2=\demi
\qw_2^{(0,-1),(1,-1)}+\demi \qw_2^{(0,1),(0,-1)}$ if $y_2=-1$
(with probability $2\alpha(1-\alpha)$), or $\tilde\qw_2=\demi
\qw_2^{(0,-1),(1,1)}+\demi \qw_2^{(0,1),(0,1)}$ if $y_2=1$ (with
probability $\alpha^2 +(1-\alpha)^2$).}
\label{fig:sg-2}
\end{center}
\end{figure}
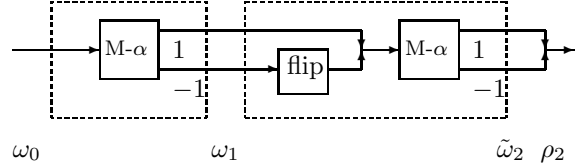
%%%%%%%%%%%%%%%%%%%%%%%%%%%%%%%%%%%%%%%%%%%%%%%%%%%%%%%%%%%%%%

These results are consistent with \cite[Section 1.3]{HW94}.
Indeed, when $\alpha=0$ (perfect measurements), the feedback
system terminates in the desired pure state $\rho_2 = \vert 1 \ra
\la 1
\vert$. The role of feedback control is clearly demonstrated here.
With imperfect measurements, $0 <
\alpha < 1$, the system terminates in the  mixed state $\rho_2$ given by
\er{eg1-rho-2},
with the degree of mixing (indicating the expected degradation in
performance) depending on the measurement error probability
parameter $\alpha$:
$$
\ba{rl}
\tr \rho_2^2 & = (\alpha^2 +\alpha(1-\alpha))^2 + (\alpha(1-\alpha)+(1-\alpha)^2)^2
\\
& < 1 \ \ \text{if} \ 0 < \alpha < 1
\\
& =1 \ \text{if} \ \alpha = 0 .
\ea
$$
\qed
\end{example}

\section{Risk-Neutral Control}
\label{sec:rn}

In this section we summarize dynamic programming results for a
well-known type of finite time horizon optimal control problem,
\cite{VPB83,KV86}. The optimal control problem discussed here can
be considered to be a prototype problem illustrating measurement
feedback in the quantum context. The dynamic programming methods
used in this paper for solving the optimal control problems are
standard, and the reader is referred to the literature for further
information, see, e.g. \cite{KA70,KV86,DB95}.

We define a {\em cost function} to be a non-negative observable
$L(u)$ that can depend on the control $u$. The cost function
encodes the designer's control objective. We also use a
non-negative observable $N$ to define a cost for the final state.

\begin{example}   \label{eg:2-d-3} (Two-state system with feedback, Example \ref{eg:2-d-2} continued.)
To set up the cost function $L(u)$ to reflect our objective of
regulating the system to the desired pure state $\vert 1 \ra$, we
define
$$
X = \frac{1}{2} \left( A- 1.I \right) = \left( \ba{cc} -1 & 0 \\ 0
& 0
\ea
\right)
$$
where $A$ is the observable corresponding to the projective
measurement \er{obs-A}. We note that the expected value of $X^2$
is
$$
\ba{rll}
 \la 1 \vert X^2 \vert 1 \ra & = \tr[ X^2 \vert 1 \ra \la 1 \vert ] & = 0
\\
\la -1 \vert X^2 \vert -1 \ra & = \tr [ X^2 \vert -1 \ra \la -1 \vert  ]  & = 1
\ea
$$
which gives zero cost to the desired state, and nonzero cost to
the undesired state. We shall also introduce a cost of control
action, as follows:
$$
c(u) = \left\{ \ba{rl} 0 & \ \text{if} \ u=0
\\
p & \ \text{if} \ u=1
\ea \right.
$$
where $p >0$. This gives zero cost for doing nothing, and a
nonzero cost for the flip operation. Thus we define the cost
function to be
\be
L(u) = X^2 + c(u)I
\label{eg1-cf}
\ee
and the cost for the final state is defined to be
$$
N=X^2 .
$$
This modifies our earlier objective of putting the system into the
desired state by including a penalty for control action.
\qed
\end{example}

Let $M > 0$ be a positive integer indicating a finite time
interval $k = 0, \ldots, M$.  Given  a sequence of control values
$u_{0,M-1} = u_0,
\ldots, u_{M-1}$ and measurements $y_{1,M} = y_{1}, \ldots, y_M$, define
the {\em risk-neutral cost functional}
\be
J_{\qw,0}(K) =  \E^K_{\qw,0} [
\sum_{i=0}^{M-1} \la
\qw_i, L(u_i) \ra + \la \qw_M, N \ra  ],% u_{0,k-1}, y_{1,k} ]
\label{rn-value-def}
\ee
where $\qw_i$, $i=0,\ldots,M$ is the solution of the system
dynamics \er{system} with initial state $\qw_0=\qw$ under the
action of a controller $K$. This is an appropriate quantum
generalization of the classical LQG cost \er{lqg}. The objective
is to minimize this functional over all measurement feedback
controllers $K \in \cK$.

Following \cite{VPB83} it is convenient to rewrite the cost
functional \er{rn-value-def}. For each $k$, given a sequence of
control values $u_{k,M-1} = u_k,
\ldots, u_{M-1}$ and measurements $y_{k+1,M} = y_{k+1}, \ldots, y_M$, define
a random sequence of observables $Q_k$ by the recursion
(\cite[equation (3.1)]{VPB83})
\be
\ba{rl}
Q_k & = \Gamma^\dagger(u_k, y_{k+1}) Q_{k+1} + L(u_k) , \ \ 0 \leq
k
\leq M-1
\\
Q_M & = N
\ea
\label{rn-cost-1}
\ee
When useful, we write
$$
Q_k = Q_k(u_{k,M-1}, y_{k+1,M})
$$
to indicate dependence on the input and outputs. $Q_k$ may be
called a {\em cost observable}. The cost functional
\er{rn-value-def} is given by
\be
\ba{l}
J_{\qw,0}(K)
\\ =\ds\sum_{y_{1,M} \in \bY^M} \la \qw, Q_0(
K(y_{1,M})_{0,M-1}, y_{1,M})
\ra
\ea
\label{rn-cost-2}
\ee
Here and elsewhere we  use abbreviations of the form
$$
K(y_{1,M})_{0,M-1} = (K_0, K_1(y_1),\ldots, K_{M-1}(y_{1,M-1}))
$$

\begin{remark} \label{rmk:rn-cost}
The cost observable $Q_k$ given by \er{rn-cost-1} and the
expression in \er{rn-cost-2} is analogous to the familiar
Heisenberg picture used in quantum physics. It is very natural
from the point of view of dynamic programming, and indeed
\er{rn-value-def} and
\er{rn-cost-2} are related by iterating
\er{rn-cost-1}. Here is the first step:
$$
\ba{rl}
\la \qw_0, Q_0 \ra & = \la \qw_0, \Gamma^\dagger(u_0,y_1) Q_1 +
L(u_0) \ra
\\
& = \la \qw_0, L(u_0) \ra + \la \Gamma(u_0,y_1)\qw_0, Q_1 \ra
\\
& = \la \qw_0, L(u_0) \ra + \la \qw_1, Q_1 \ra p(y_1\vert u_0,
\qw_0)
\ea
$$
where $\qw_1=\Lambda_\Gamma(u_0,y_1)\qw_0$ and $p(y_1\vert u_0,
\qw_0)$ is given by \er{trans-prob}.
\qed
\end{remark}

The key idea of dynamic programming is to look at the current
state at a current time $0\leq k\leq M-1$ and to optimize the
remaining cost from the current time to the final time. This leads
to an iterative solution. Accordingly, we define, for each $0 \leq
k
\leq M$, the cost to go incurred by a controller $K$ (restricted
to $k
\leq l
\leq M-1$) to be
\be
\ba{l}
J_{\qw,k}(K)
\\ =  \ds\sum_{y_{k+1,M} \in \bY^{M-k}} \la
\qw, Q_k( K(y_{k+1,M})_{k,M-1}, y_{k+1,M})
\ra
\ea
\label{rn-value}
\ee
The dynamic programming equation associated with this risk-neutral
problem is
\be
\ba{rl}
V(\qw,k) &= \ds\inf_{u \in \bU} \{ \la \qw, L(u) \ra
\\& \ + \ds\sum_{y
\in \bY} V(
\Lambda_\Gamma (u,y)\qw , k+1) p(y
\vert u,
\qw) \},
\\
V(\qw,M) & = \la \qw, N \ra
\ea
\label{rn-dpe}
\ee
where $0 \leq k \leq M-1$. This is the fundamental equation from
which optimality or otherwise of a controller can be determined.

Let $V$ be the solution to the dynamic programming equation
\er{rn-dpe}. Then for any controller $K \in \cK$ we have
\be
V(\qw,k) \leq J_{\qw,k}(K) .
\label{VJ-1}
\ee
If we assume in addition that a minimizer
\be
\ba{rl}
\bu^\ast(\qw,k) & \in \ds\argmin_{u \in \bU} \{  \la \qw, L(u) \ra
\\ &+ \ds\sum_{y\in \bY}  V(
\Lambda_\Gamma (u,y)\qw , k+1) p(y
\vert u,
\qw)  ) \}
\ea
\label{rn-u-star}
\ee
exists\footnote{The notation $\argmin_{u\in\bU} f(u)$ means the
subset of values from $\bU$ minimizing $f$.}
 for all $\qw$, $0 \leq k \leq M-1$, then the separation
structure controller $K^{\bu^\ast}_{\qw_0}$ (recall section
\ref{sec:cd}) defined by
\er{rn-u-star} is optimal, i.e.
\be
J_{\qw_0,0}(K^{\bu^\ast}_{\qw_0}) = V(\qw_0,0) \leq J_{\qw_0,0}(K)
\label{VJ-2}
\ee
for all $K \in \cK$.

\begin{example}   \label{eg:2-d-4} (Two-state system with feedback, Example \ref{eg:2-d-3} continued.)
We solve the dynamic programming equation \er{rn-dpe} and
determine the optimal feedback controls as follows. For $k=M=2$ we
have
$$
V(\qw, 2) = \la \qw, X^2 \ra  = \qw_{11}
$$
and hence for $k=1$
$$
V(\qw,1) = \qw_{11} + \min[V_0(\qw,1), V_1(\qw,1)]
$$
where where $V_0(\qw,1), V_1(\qw,1)$ are given in Appendix
\ref{app:sg-eg}. Hence we obtain
$$
\bu^{\ast}(\qw,1) = \left\{  \ba{rl}
0 & \ \text{if} \ V_0(\qw,1) \leq V_1(\qw,1)
\\
1 & \ \text{if} \ V_0(\qw,1) > V_1(\qw,1) .
 \ea \right.
$$
At time $k=0$ we have
$$
V(\qw,0) = \qw_{11} + \min[V_0(\qw,0), V_1(\qw,0)]
$$
where $V_0(\qw,0), V_1(\qw,0)$ are given in Appendix
\ref{app:sg-eg}, which gives
$$
\bu^\ast(\qw,0) = \left\{  \ba{rl}
0 & \ \text{if} \ V_0(\qw,0) \leq V_1(\qw,0)
\\
1 & \ \text{if} \ V_0(\qw,0) > V_1(\qw,0) .
 \ea \right.
$$

The optimal risk-neutral feedback controller is given by
$$
u_0=K^{\bu^\ast}_{\qw_0,0} = \bu^\ast(\qw_0,0), \ \
u_1=K^{\bu^\ast}_{\qw_0,1}(y_1) =
\bu^\ast(
\qw_1,1)
$$
where $\qw_1 = \Lambda_\Gamma(u_0,y_1)\qw_0$. {\em Note that the
control $u_1$ depends on $y_1$ through the conditional state
$\qw_1$ (separation structure).} A physical implementation of the
quantum system with optimal risk-neutral feedback is shown in
Figure
\ref{fig:sg-2-rn}.

%%%%%%%%%%%%%%%%%%%%%%%%%%%%%%%%%%%%%%%%%%%%%%%%%%%%%%%%%%%%%%
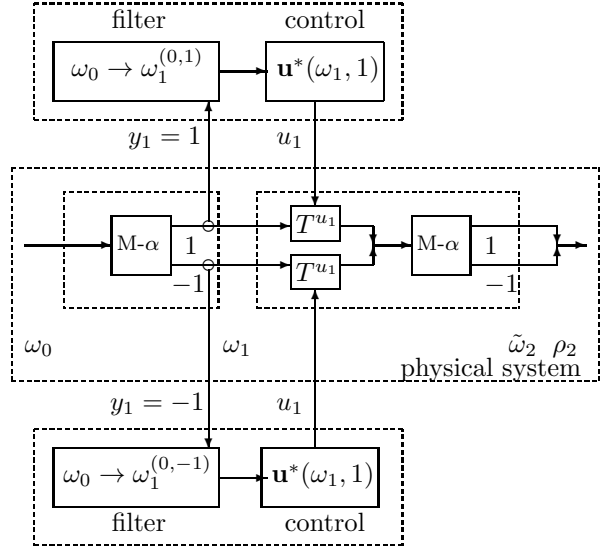
\begin{figure}[htb]
\begin{center}
\setlength{\unitlength}{1600sp}%
\begingroup\makeatletter\ifx\SetFigFont\undefined%
\gdef\SetFigFont#1#2#3#4#5{%
  \reset@font\fontsize{#1}{#2pt}%
  \fontfamily{#3}\fontseries{#4}\fontshape{#5}%
  \selectfont}%
\fi\endgroup%
\begin{picture}(9324,8424)(2089,-6973)         %\begin{picture}(9324,8424)(2089,-6973)
{\color[rgb]{0,0,0}\thinlines
\put(5251,-2611){\circle{150}}
}%
{\color[rgb]{0,0,0}\put(5251,-2011){\circle{150}}
}%
{\color[rgb]{0,0,0}\put(8401,-2761){\framebox(900,900){}}
}%
{\color[rgb]{0,0,0}\put(3751,-2761){\framebox(900,900){}}
}%
{\color[rgb]{0,0,0}}%
{\color[rgb]{0,0,0}}%
{\color[rgb]{0,0,0}\put(7801,-2311){\vector( 1, 0){600}}
}%
{\color[rgb]{0,0,0}\put(2401,-2311){\vector( 1, 0){1350}}
}%
{\color[rgb]{0,0,0}}%
{\color[rgb]{0,0,0}\put(9301,-2611){\line( 1, 0){1350}}
\put(10651,-2611){\vector( 0, 1){300}}
}%
{\color[rgb]{0,0,0}\put(9301,-2011){\line( 1, 0){1350}}
\put(10651,-2011){\vector( 0,-1){300}}
}%
{\color[rgb]{0,0,0}\put(10651,-2311){\vector( 1, 0){450}}
}%
{\color[rgb]{0,0,0}}%
{\color[rgb]{0,0,0}\put(6526,-2236){\framebox(750,525){}}
}%
{\color[rgb]{0,0,0}\put(6526,-2986){\framebox(750,525){}}
}%
{\color[rgb]{0,0,0}\put(4651,-2011){\vector( 1, 0){1875}}
}%
{\color[rgb]{0,0,0}\put(4651,-2611){\vector( 1, 0){1875}}
}%
{\color[rgb]{0,0,0}\put(7276,-2611){\line( 1, 0){525}}
\put(7801,-2611){\vector( 0, 1){300}}
}%
{\color[rgb]{0,0,0}\put(7276,-2011){\line( 1, 0){525}}
\put(7801,-2011){\vector( 0,-1){300}}
}%
{\color[rgb]{0,0,0}\put(6001,-3286){\dashbox{60}(4050,1800){}}
}%
{\color[rgb]{0,0,0}\put(3001,-3286){\dashbox{60}(2400,1800){}}
}%
{\color[rgb]{0,0,0}\put(2201,-4411){\dashbox{60}(9100,3300){}} %60,9300,3300
}%
{\color[rgb]{0,0,0}\put(5401,-5911){\vector( 1, 0){750}}
}%
{\color[rgb]{0,0,0}\put(5251,-2686){\vector( 0,-1){2775}}
}%
{\color[rgb]{0,0,0}\put(6901,-5461){\vector( 0, 1){2475}}
}%
{\color[rgb]{0,0,0}\put(5401,389){\vector( 1, 0){750}}
}%
{\color[rgb]{0,0,0}\put(5251,-1936){\vector( 0, 1){1875}}
}%
{\color[rgb]{0,0,0}\put(6901,-61){\vector( 0,-1){1650}}
}%
{\color[rgb]{0,0,0}\put(2851,-61){\framebox(2550,900){}}
}%
{\color[rgb]{0,0,0}\put(6151,-61){\framebox(1800,900){}}
}%
{\color[rgb]{0,0,0}\put(2551,-361){\dashbox{60}(5700,1800){}}
}%
{\color[rgb]{0,0,0}\put(2851,-6361){\framebox(2550,900){}}
}%
{\color[rgb]{0,0,0}\put(6076,-6361){\framebox(1800,900){}}
}%
{\color[rgb]{0,0,0}\put(2551,-6961){\dashbox{60}(5700,1800){}}
}%
\put(8476,-2386){\makebox(0,0)[lb]{\smash{\SetFigFont{8}{14.4}{\familydefault}{\mddefault}{\updefault}{\color[rgb]{0,0,0}M-$\alpha$}%
}}}
\put(3826,-2386){\makebox(0,0)[lb]{\smash{\SetFigFont{8}{14.4}{\familydefault}{\mddefault}{\updefault}{\color[rgb]{0,0,0}M-$\alpha$}%
}}}
\put(4876,-2436){\makebox(0,0)[lb]{\smash{\SetFigFont{10}{14.4}{\familydefault}{\mddefault}{\updefault}{\color[rgb]{0,0,0}$1$}%
}}}
\put(9526,-2436){\makebox(0,0)[lb]{\smash{\SetFigFont{10}{14.4}{\familydefault}{\mddefault}{\updefault}{\color[rgb]{0,0,0}$1$}%
}}}
\put(4676,-3036){\makebox(0,0)[lb]{\smash{\SetFigFont{10}{14.4}{\familydefault}{\mddefault}{\updefault}{\color[rgb]{0,0,0}$-1$}%
}}}
\put(9526,-3036){\makebox(0,0)[lb]{\smash{\SetFigFont{10}{14.4}{\familydefault}{\mddefault}{\updefault}{\color[rgb]{0,0,0}$-1$}%
}}}
\put(2401,-3961){\makebox(0,0)[lb]{\smash{\SetFigFont{10}{14.4}{\familydefault}{\mddefault}{\updefault}{\color[rgb]{0,0,0}$\qw_0$}%
}}}
\put(10576,-3961){\makebox(0,0)[lb]{\smash{\SetFigFont{10}{14.4}{\familydefault}{\mddefault}{\updefault}{\color[rgb]{0,0,0}$\rho_2$}%
}}}
\put(9901,-3961){\makebox(0,0)[lb]{\smash{\SetFigFont{10}{14.4}{\familydefault}{\mddefault}{\updefault}{\color[rgb]{0,0,0}$\tilde\qw_2$}%
}}}
\put(5476,-3961){\makebox(0,0)[lb]{\smash{\SetFigFont{10}{14.4}{\familydefault}{\mddefault}{\updefault}{\color[rgb]{0,0,0}$\qw_1$}%
}}}
\put(6601,-2111){\makebox(0,0)[lb]{\smash{\SetFigFont{10}{14.4}{\familydefault}{\mddefault}{\updefault}{\color[rgb]{0,0,0}$T^{u_1}$}%
}}}
\put(6601,-2861){\makebox(0,0)[lb]{\smash{\SetFigFont{10}{14.4}{\familydefault}{\mddefault}{\updefault}{\color[rgb]{0,0,0}$T^{u_1}$}%
}}}
\put(8226,-4336){\makebox(0,0)[lb]{\smash{\SetFigFont{10}{14.4}{\familydefault}{\mddefault}{\updefault}{\color[rgb]{0,0,0}physical system}%
}}}
\put(6226,-5986){\makebox(0,0)[lb]{\smash{\SetFigFont{10}{14.4}{\familydefault}{\mddefault}{\updefault}{\color[rgb]{0,0,0}$\bu^\ast(\qw_1,1)$}%
}}}
\put(3001,-5986){\makebox(0,0)[lb]{\smash{\SetFigFont{10}{14.4}{\familydefault}{\mddefault}{\updefault}{\color[rgb]{0,0,0}$\qw_0 \to\qw_1^{(0,-1)}$}%
}}}
\put(6301,314){\makebox(0,0)[lb]{\smash{\SetFigFont{10}{14.4}{\familydefault}{\mddefault}{\updefault}{\color[rgb]{0,0,0}$\bu^\ast(\qw_1,1)$}%
}}}
\put(3126,314){\makebox(0,0)[lb]{\smash{\SetFigFont{10}{14.4}{\familydefault}{\mddefault}{\updefault}{\color[rgb]{0,0,0}$\qw_0\to\qw_1^{(0,1)}$}%
}}}
\put(3751,1064){\makebox(0,0)[lb]{\smash{\SetFigFont{10}{14.4}{\familydefault}{\mddefault}{\updefault}{\color[rgb]{0,0,0}filter}%
}}}
\put(6426,1064){\makebox(0,0)[lb]{\smash{\SetFigFont{10}{14.4}{\familydefault}{\mddefault}{\updefault}{\color[rgb]{0,0,0}control}%
}}}
\put(4001,-736){\makebox(0,0)[lb]{\smash{\SetFigFont{10}{14.4}{\familydefault}{\mddefault}{\updefault}{\color[rgb]{0,0,0}$y_1=1$}%
}}}
\put(6301,-736){\makebox(0,0)[lb]{\smash{\SetFigFont{10}{14.4}{\familydefault}{\mddefault}{\updefault}{\color[rgb]{0,0,0}$u_1$}%
}}}
\put(6301,-4861){\makebox(0,0)[lb]{\smash{\SetFigFont{10}{14.4}{\familydefault}{\mddefault}{\updefault}{\color[rgb]{0,0,0}$u_1$}%
}}}
\put(3726,-4861){\makebox(0,0)[lb]{\smash{\SetFigFont{10}{14.4}{\familydefault}{\mddefault}{\updefault}{\color[rgb]{0,0,0}$y_1=-1$}%
}}}
\put(3751,-6736){\makebox(0,0)[lb]{\smash{\SetFigFont{10}{14.4}{\familydefault}{\mddefault}{\updefault}{\color[rgb]{0,0,0}filter}%
}}}
\put(6426,-6736){\makebox(0,0)[lb]{\smash{\SetFigFont{10}{14.4}{\familydefault}{\mddefault}{\updefault}{\color[rgb]{0,0,0}control}%
}}}
\end{picture}

\caption{Physical realization of the two stages of the two-state system with feedback using the optimal
risk-neutral controller $K^{\bu^\ast}_{\qw_0}$ (with $\qw_0$ given
by
\er{eg1-w0}, we have  $u_0=\bu^\ast(\qw_0,0)=0$, $u_1=\bu^\ast(\qw_1,1)$).}
\label{fig:sg-2-rn}
\end{center}
\end{figure}
%%%%%%%%%%%%%%%%%%%%%%%%%%%%%%%%%%%%%%%%%%%%%%%%%%%%%%%%%%%%%%

Let's consider the special case $\alpha = 0$ and $p=0$, with
initial state \er{eg1-w0}. We then find that
$V_0(\qw_0,0)=V_1(\qw_0,0)=0.5$, and hence we take
$\bu^\ast(\qw_0,0)=0$; i.e. $u_0=0$.

Next, if $y_1=-1$ is observed, we have $\qw_1 = \vert -1 \ra \la
-1 \vert$, $V_0(\qw_1,1)=1$ and $V_1(\qw_1,1)= 0$. Hence we take
$\bu^\ast(\qw_1,1)= 1$, i.e. $u_1=1$. However, if $y_1=1$ is
observed, we have $\qw_1 = \vert 1 \ra
\la 1 \vert$, $V_0(\qw_1,1)=0$ and $V_1(\qw_1,1)= 1$; and hence we
take $\bu^\ast(\qw_1,1)= 0$, i.e. $u_1=0$. In either case we
achieve the desired state $\rho_2=\qw_2=\vert 1 \ra\la 1 \vert$.

This action is the same as that seen before for the controller
$\bar K$. The same controller is obtained for $0 < \alpha < 0.5$
and $p=0$, but $\qw_2$ will be a mixed state. If $p\neq 0$ the
optimal controller $K^{\bu^\ast}_{\qw_0}$ will result in control
actions that in general differ from those of $\bar K$.
\qed
\end{example}

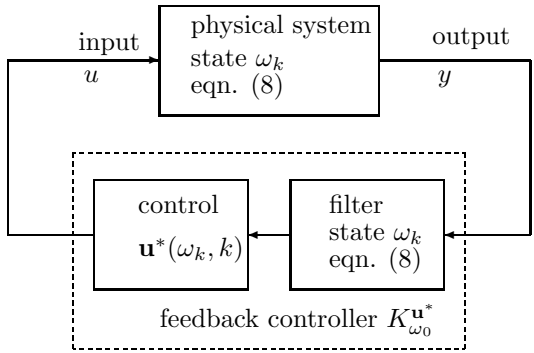
\begin{figure}[htb]
\begin{center}
\setlength{\unitlength}{1800sp}%
\begingroup\makeatletter\ifx\SetFigFont\undefined%
\gdef\SetFigFont#1#2#3#4#5{%
  \reset@font\fontsize{#1}{#2pt}%
  \fontfamily{#3}\fontseries{#4}\fontshape{#5}%
  \selectfont}%
\fi\endgroup%
\begin{picture}(7224,4749)(889,-4798)
\thinlines
{\color[rgb]{0,0,0}\put(3001,-1561){\framebox(3000,1500){}}
}%
{\color[rgb]{0,0,0}\put(2101,-3961){\framebox(2100,1500){}}
}%
{\color[rgb]{0,0,0}\put(4801,-3961){\framebox(2100,1500){}}
}%
{\color[rgb]{0,0,0}\put(4801,-3211){\vector(-1, 0){600}}
}%
{\color[rgb]{0,0,0}\put(2101,-3211){\line(-1, 0){1200}}
\put(901,-3211){\line( 0, 1){2400}}
\put(901,-811){\vector( 1, 0){2100}}
}%
{\color[rgb]{0,0,0}\put(6001,-811){\line( 1, 0){2100}}
\put(8101,-811){\line( 0,-1){2400}}
\put(8101,-3211){\vector(-1, 0){1200}}
}%
{\color[rgb]{0,0,0}\put(1801,-4786){\dashbox{60}(5400,2700){}}
}%
\put(1951,-1111){\makebox(0,0)[lb]{\smash{\SetFigFont{10}{14.4}{\familydefault}{\mddefault}{\updefault}{\color[rgb]{0,0,0}$u$}%
}}}
\put(6826,-1111){\makebox(0,0)[lb]{\smash{\SetFigFont{10}{14.4}{\familydefault}{\mddefault}{\updefault}{\color[rgb]{0,0,0}$y$}%
}}}
\put(1876,-661){\makebox(0,0)[lb]{\smash{\SetFigFont{10}{14.4}{\familydefault}{\mddefault}{\updefault}{\color[rgb]{0,0,0}input}%
}}}
\put(6751,-586){\makebox(0,0)[lb]{\smash{\SetFigFont{10}{14.4}{\familydefault}{\mddefault}{\updefault}{\color[rgb]{0,0,0}output}%
}}}
\put(3426,-1261){\makebox(0,0)[lb]{\smash{\SetFigFont{10}{14.4}{\familydefault}{\mddefault}{\updefault}{\color[rgb]{0,0,0}eqn. \er{system}}%
}}}
\put(3426,-886){\makebox(0,0)[lb]{\smash{\SetFigFont{10}{14.4}{\familydefault}{\mddefault}{\updefault}{\color[rgb]{0,0,0}state $\qw_k$}%
}}}
\put(3426,-436){\makebox(0,0)[lb]{\smash{\SetFigFont{10}{14.4}{\familydefault}{\mddefault}{\updefault}{\color[rgb]{0,0,0}physical system}%
}}}
\put(2701,-2911){\makebox(0,0)[lb]{\smash{\SetFigFont{10}{14.4}{\familydefault}{\mddefault}{\updefault}{\color[rgb]{0,0,0}control}%
}}}
\put(2701,-3511){\makebox(0,0)[lb]{\smash{\SetFigFont{10}{14.4}{\familydefault}{\mddefault}{\updefault}{\color[rgb]{0,0,0}$\bu^\ast(\qw_k,k)$}%
}}}
\put(5326,-2911){\makebox(0,0)[lb]{\smash{\SetFigFont{10}{14.4}{\familydefault}{\mddefault}{\updefault}{\color[rgb]{0,0,0}filter}%
}}}
\put(5326,-3286){\makebox(0,0)[lb]{\smash{\SetFigFont{10}{14.4}{\familydefault}{\mddefault}{\updefault}{\color[rgb]{0,0,0}state $\qw_k$}%
}}}
\put(5326,-3661){\makebox(0,0)[lb]{\smash{\SetFigFont{10}{14.4}{\familydefault}{\mddefault}{\updefault}{\color[rgb]{0,0,0}eqn. \er{system}}%
}}}
\put(3001,-4486){\makebox(0,0)[lb]{\smash{\SetFigFont{10}{14.4}{\familydefault}{\mddefault}{\updefault}{\color[rgb]{0,0,0}feedback controller $K^{\bu^\ast}_{\qw_0}$}%
}}}
\end{picture}

\caption{Optimal risk-neutral controller $K^{\bu^\ast}_{\qw_0}$ showing separation structure and states of the
physical system $\qw_k$ and filter $\qw_k$.}
\label{fig:fb-rn}
\end{center}
\end{figure}

\begin{remark}   \label{rmk:rn-fb}
Note that the optimal risk-neutral controller
$K^{\bu^\ast}_{\qw_0}$ determined by
\er{rn-u-star} feeds back the conditional state $\qw_k$, given by the
SME \er{system}, in accordance with its separation structure,
Figure \ref{fig:fb-rn}. We note that the conditional state $\qw_k$
obtained from
\er{system} provides the optimal means for calculating estimates
of observables (in the sense of minimum mean square error), viz.
$\la
\qw_k, B \ra$, and can be regarded as the optimal filter in this sense. This
means that from the point of view of optimal risk-neutral control,
the best thing to do is to make use of the optimal filter
\er{system}, a dynamical quantity that contains {\em knowledge}
of the quantum system, as obtained by the controller through the
measurement process embedded in $\Gamma$.
\qed
\end{remark}

%%%%%%%%%%%%%%%%%%%%%%%%%%%%%%%%%%%%%%%%%%%%%%%%%%%%%%%%%%%%%%
%%%%%%%%%%%%%%%%%%%%%%%%%%%%%%%%%%%%%%%%%%%%%%%%%%%%%%%%%%%%%%

\begin{remark}    \label{rmk:dual}
A second remark we wish to make here concerns the well-known
concept in control engineering of {\em dual control},
\cite[Chapter 6.8]{KV86}. This concept relates to the dual function of the
measurement feedback controller $K$, viz. (i) to alter the future
evolution of the system, and (ii) to alter the future values of
the available information. The optimal choice of $K$ takes both of
these factors into account.
\qed
\end{remark}

\begin{remark}    \label{rmk:robust-1}
The final remark for this section concerns {\em feedback} and {\em
robustness}. Feedback is the most important concept in control
engineering, and has a long history going back at least to the
mechanical governors used to regulate the speed of steam engines.
Feedback is used to compensate for disturbances and uncertainty,
and feedback loops typically enjoy a robustness margin (e.g. gain
margin and phase margin in classical control engineering), a
measure of this compensation ability. Note that in the absence of
disturbances and uncertainty, feedback is completely unnecessary
and control can be achieved by a prescribed open loop controller.
However, in reality both quantum and classical systems are subject
to disturbances and uncertainty, e.g. (i) the influence of an
environment, (ii) model error due to approximation and unknown
parameters, and (iii) imprecise measurements. In the quantum
context, there is the further complication \cite{DJJ00} that as a
consequence of the act of measurement randomness is introduced,
and this could potentially reduce control effectiveness.
Measurement feedback control of quantum systems is fundamentally a
stochastic control problem containing non-classical
characteristics. These considerations underscore the importance of
feedback control and the need for robustness when controlling
quantum systems. Robustness issues will be taken up again in
section
\ref{sec:rs} (see Example
\ref{eg:simple}).
\qed
\end{remark}

\section{Multiplicative Costs and Risk-Sensitive Control}
\label{sec:rs}

We turn now to the risk-sensitive optimal control problem, the
main object of this paper. The {\em risk-sensitive cost
functional} we consider, a quantum generalization of LEQG
\er{leqg}, is
\be
J^{\mu}_{\qw,0}(K) \\
   = \E_{\qw,0}[\prod_{k=0}^{M-1} \la
\qw_k,e^{\mu L(u_k)}\ra
\la \qw_M, e^{\mu N} \ra ]
\label{r-rs-1}
\ee
where $\mu > 0$ is a positive risk parameter, $L(u)$ is a cost
function (as defined in section \ref{sec:rn}), and $N$ is a
non-negative observable. The conditional states $\qw_k$ are given
by the quantum system model
\er{system}.

\begin{remark} \label{rmk:rs-cost}
The risk-sensitive cost \er{r-rs-1}, by use of the exponential
function, gives heavy weight to large values of the cost functions
in the exponents. A system controlled by a controller minimizing
this cost is not likely to experience large values of these
quantities. Risk-sensitive controllers are known to enjoy some
robustness properties against uncertainty in the model and
external disturbances, see
\cite{DJP00} and Example \ref{eg:simple}.
\qed
\end{remark}

The primary goal in this section is to find the optimal controller
for the risk-sensitive cost functional \er{r-rs-1}. As noted, this
cost functional is defined in terms of the conditional state
$\qw_k$ of the quantum system \er{system}. However, in order to
solve this optimization problem, we need to express the cost
functional in a manner that facilitates the use of optimal control
methods. As in the classical LEQG case, this requires the
introduction of a new state, which in general is {\em
unnormalized}. To define this new unnormalized state, for which we
use the notation $\qwr$ to distinguish such states from normalized
states $\qw$, we need to use possibly nonlinear operators
(observables) $B$ and (super) operators $R$. These nonlinear
operators allow us to formulate and solve a general class of
multiplicative cost optimal control problems for quantum
systems\footnote{We denote the value of $B$ at $\qwr$ by $\la
\qwr, B \ra$, extending the notation \er{tr-ip}. The (generalized)
adjoint $R^\dagger$ of $R$ is defined by $\la
\qwr, R^\dagger B
\ra = \la R\qwr, B\ra$ for all $\qwr$ and all $B$.}.

Our risk-sensitive and multiplicative cost functionals can be
defined in terms of  {\em (super) operator valued costs} $R(u)$
that satisfy the real multiplicative homogeneity property
\be
R(u) r \qwr =  r R(u)\qwr
\label{mcf-h}
\ee
for any real number $r$ and any $\qwr$, $u$. The risk-sensitive
problem corresponds to particular choices of operator valued cost,
Example
\ref{eg:mcf-rs}. However, the fundamental equations in this
section are valid for any operator valued cost $R(u)$ satisfying
\er{mcf-h}. Note that operator valued costs $R(u)$ are not in general
quantum operations (because linearity and the inequality $\la
R(u)\qwr,I\ra
\leq
\la
\qwr,I\ra$ need not hold in general).

\begin{example} \label{eg:mcf-rs}
We give two examples of  operator valued costs.

(i) A specific linear form for $R(u)$ is
\be
R(u) \qwr = \sum_c Z_c(u) \qwr Z_c(u)
\label{mcf-sum}
\ee
where $\{ Z_c(u) \}$ is a family of cost functions (section
\ref{sec:rn}).
The adjoint $R^\dagger(u)$ acts on observables $B$ via
\be
R^\dagger(u) B = \sum_c Z_c(u) B Z_c(u) ,
\label{mcf-sum-a}
\ee
and thereby defines a linear functional on unnormalized states by
\be
\la \qwr, R^\dagger(u) B \ra = \sum_c \la Z_c(u) \qwr Z_c(u) , B
\ra
\label{mcf-sum-a-2}
\ee
(we have written this explicitly to facilitate comparison with
\er{r-rs-2a} below).

(ii) An operator valued cost $R(u)$ corresponding to the
risk-sensitive cost \er{r-rs-1} can be defined as follows. Let
$L(u)$ be a cost function, and $\mu > 0$. Then set
\be
R(u)\qwr = \frac{\la \qwr, e^{\mu L(u)} \ra}{\la \qwr,1\ra} \qwr .
\label{r-rs-2}
\ee
Note that $R(u)$ is nonlinear, but satisfies the real
multiplicative homogeneity condition \er{mcf-h}. The adjoint
operator
$R^\dagger(u)$ applied to an operator $B$ is a nonlinear
functional of $\qwr$ given by
\be
\la \qwr, R^\dagger(u)B \ra = \frac{\la \qwr, e^{\mu L(u)} \ra}{\la \qwr,1\ra} \la
\qwr, B \ra
\label{r-rs-2a}
\ee
(cf. \er{mcf-sum-a-2} above).

The relationship between $R(u)$ and the risk-sensitive cost
\er{r-rs-1} will be explained in Example \ref{eg:mcf-rs-ru} below.

\qed
\end{example}

Given an operator valued cost $R(u)$, we shall find it convenient
to introduce an operator $\Gamma_R(u,y)$ defined by
\be
\Gamma_R(u,y)=\Gamma(u,y)R(u) .
\label{gamma-R}
\ee
In general, $\Gamma_R$ is  not normalized:
$$
\sum_{y \in \bY}  \la \Gamma_R(u,y)\qwr, I \ra =  \la R(u)\qwr, I \ra
\neq \la \qwr, I \ra .
$$
The operator $\Gamma_R$ will be used to define a new state
evolution as follows. Define an operator $\Lambda_{\Gamma,R}$ by
\be
 \Lambda_{\Gamma,R}(u,y)\qwr =
\ds
\frac{\Gamma_R(u,y)\qwr}{ p_R(y \vert u, \qwr)}
\label{gamma-dagger-bar-La}
\ee
where
\be
p_R(y \vert u, \qwr) = \ds\frac{\la \Gamma_R(u,y)\qwr, I
\ra}{\la R(u)\qwr,I \ra} .
\label{trans-prob-L}
\ee
In general, the state $\Lambda_{\Gamma,R}(u,y)\qwr$ is {\em
unnormalized}. However, $p_R(y\vert u,\qwr)$ is a probability
distribution, since it is easy to check that
$$
\sum_{y \in
\bY}p_R(y
\vert u,
\qwr) = 1.
$$
However, we point out that
\be
\la \Lambda_{\Gamma, R}(u,y)\qwr,I \ra = \la R(u)\qwr, I
\ra .
\label{lambda-gamma-l-unnorm}
\ee
This unnormalized state transition operator arises  in the dynamic
programming equation, as we shall see below.

Associated with the operator $\Lambda_{\Gamma, R}$ is the dynamics
\be
 \qwr_{k+1} = \Lambda_{\Gamma,R}(u_k, y_{k+1})  \qwr_k,
\label{rs-is-dyn}
\ee
where $y_{k+1}$ is distributed according to the probability
distribution $p_R( y_{k+1} \vert u_k, \qwr_k)$ given  by
\er{trans-prob-L}. This is a controlled Markov chain, with
{\em unnormalized} states $\qwr_k$. It is a modified stochastic
master equation corresponding to the operator $\Gamma_R$. Under
the action of a controller $K
\in \cK$ the stochastic process $\qwr_k$ is determined by
\er{rs-is-dyn} and $u_k = K_k(y_{1,k})$.

The separation structure controller in this case takes the
following form. Given a function $\hat\bu(\qwr,k)$ and initial
state $\qwr_0$ we define a controller $K^{\hat\bu}_{\qwr_0}
\in
\cK$ by
$$
u_k = \hat\bu(\qwr_k,k)
$$
where $\qwr_k$ is given by \er{rs-is-dyn}, $0 \leq k \leq M$, with
initial condition $\qwr_0$.

Let $M > 0$ be a positive integer indicating a finite time
interval $k = 0, \ldots, M$. For each $k$, given  a sequence of
control values $u_{k,M-1} = u_k,
\ldots, u_{m-1}$ and measurement values $y_{k+1,M} = y_{k+1}, \ldots, y_M$, define
random cost observables $G_k$ by the recursion
\be
\ba{rl}
G_k & = R^\dagger(u_k) \Gamma^\dagger(u_k, y_{k+1}) G_{k+1} ,
\\
G_M & = F
\ea
\label{rs-cost-1}
\ee
where $0 \leq k \leq M-1$ and $F$ is a non-negative linear
observable. It is evident that $G_k$ is real multiplicative
homogeneous if
$G_{k+1}$ is (recall \er{mcf-h}).

We next define the {\em multiplicative} cost functional
\be
J^\mu_{\qwr,0}(K) =\sum_{y_{1,M} \in \bY^M} \la \qwr, G_0(
K(y_{1,M})_{0,M-1}, y_{1,M}) \ra
\label{rs-cost-2}
\ee
where $K \in \cK$ is a measurement feedback controller.

\begin{lemma} \label{lemma:rs-alt}
The cost functional $J^\mu_{\qwr,0}(K)$ defined by \er{rs-cost-2}
is given by the alternate expression
\be
J^\mu_{\qwr,0}(K) =  \E_{\qwr,0}^K[ \la \qwr_M,F
\ra ]
\label{rs-value-alt}
\ee
where $\qwr_i$, $i=k,\ldots,M$ is the solution of the recursion
\er{rs-is-dyn} with initial state $\qwr_0=\qwr$ under the action of
the controller $K$.
\end{lemma}

\begin{proof}
We have
$$
\ba{rl}
\la \qwr_0, G_0 \ra & = \la \qwr_0, R^\dagger(u_0)
\Gamma^\dagger(u_0,y_1) G_1\ra
\\
& = \la
 R(u_0)\qwr_0,\Gamma(u_0,y_1)^\dagger G_1 \ra
\\
&= \la
\Gamma(u_0,y_1) R(u_0)\qwr_0,G_1 \ra
\\
& = \la \qwr_1, G_1) \ra p_R(y_1 \vert u_0,\qwr_0)
\ea
$$
where $\qwr_1 = \Lambda_{\Gamma,R}(u_0,y_1)\qwr_0$ and $p_R(y_1
\vert u_0,\qwr_0)$ is given by \er{trans-prob-L}. Iterating in
this way we see that \er{rs-cost-2} and \er{rs-value-alt} are
equivalent. These properties use the real multiplicative
homogeneity property of $G_k$.
\qed
\end{proof}

\begin{example}  \label{eg:mcf-rs-ru}
(Continuation of Example \ref{eg:mcf-rs} (ii).) We now show that
when $R(u)$ is given by \er{r-rs-2}, and
$$
F=e^{\mu N},
$$
where $N$ is a non-negative linear observable, the multiplicative
cost functional
$J^\mu_{\qwr,0}(K)$ defined by
\er{rs-cost-2} equals the risk-sensitive cost functional
\er{r-rs-1}.

Proceeding as in the proof of Lemma \ref{lemma:rs-alt} we have
$$
\ba{rl}
& \la \qwr_0, G_0 \ra
\\
& = \la \qwr_0, R^\dagger(u_0)
\Gamma^\dagger(u_0,y_1) G_1\ra
\\
&= \la
\Gamma(u_0,y_1) R(u_0)\qwr_0,G_1 \ra
\\
& =  \la \Gamma(u_0,y_1) \qwr_0,G_1 \ra \ds{\frac{\la \qwr_0,
e^{\mu L(u_0)}\ra}{\la \qwr_0,1 \ra} }
\\
& =
\ds{\frac{\la \Gamma(u_0,y_1)\qwr_0, G_1 \ra/\la \qwr_0,1  \ra  }{\la \Gamma(u_0,y_1)\qwr_0,1 \ra/\la \qwr_0,1  \ra  }
\la \qwr_0, e^{\mu L(u_0)} \ra .
}
\\
& \hspace{3.0cm} . \ds{\frac{\la \Gamma(u_0,y_1)\qwr_0,1
\ra}{\la
\qwr_0,1 \ra}}
\\
& =
\la \Lambda_\Gamma(u_0,y_1)\bar\qwr_0, G_1 \ra \la \qwr_0, e^{\mu L(u_0)} \ra p(y_1 \vert u_0,
\bar\qwr_0)
\ea
$$
where $\bar\qwr_0 = \qwr_0/\la \qwr_0,1  \ra$,
$\Lambda_\Gamma(u,y)$ is defined by \er{gamma-dagger-bar}   and
$p(y\vert u,\qw)$ is defined by \er{trans-prob}.  Now if $\qwr_0=\qw_0$ is normalized, with
$\la \qw_0,1\ra=1$, then we have shown that
$$
\ba{rl}
& \la \qw_0, G_0\ra
\\
& =\la \Lambda_\Gamma(u_0,y_1)\qw_0, G_1 \ra
\la \qw_0, e^{\mu L(u_0)} \ra p(y_1 \vert u_0,
\bar\qw_0)
\\
& =\la \qw_1, G_1 \ra \la \qw_0, e^{\mu L(u_0)} \ra p(y_1 \vert
u_0,
\qw_0)
\ea
$$
where $\qw_1 = \Lambda_\Gamma(u_0,y_1)\qw_0$ is the normalized
state evolving according to the quantum system model \er{system}.
Note that $\la
\qw_1,1\ra=1$.  Continuing in this way we see that \er{rs-cost-2}
equals the risk-sensitive cost functional \er{r-rs-1}, using Lemma
\ref{lemma:rs-alt}.

It can also be checked that $\qw_k$ and $\qwr_k$ are related
simply via
\be
\qwr_k = \prod_{i=0}^{k-1} \la \qw_i, e^{\mu L(u_i)} \ra  \qw_k .
\label{simple}
\ee

\qed
\end{example}

To solve the optimal control problem for the cost functional
\er{rs-cost-2}, we define the cost to go
\be
\ba{l}
J^\mu_{\qwr,k}(K)
\\=  \ds\sum_{ y_{k+1,M} \in \bY^{M-k}}
\la
\qwr, G_k( K(y_{k+1,M})_{k,M-1}, y_{k+1,M})
\ra
\ea
\label{rs-value}
\ee
and the corresponding dynamic programming equation
\be
\ba{rl}
W(\qwr,k) &= \ds\inf_{u \in \bU} \{ \sum_{y \in \bY} W(
\Lambda_{\Gamma, R} (u,y)  \qwr , k+1)
\\
& \hspace{2.0cm} .p_R(y
\vert u,
\qwr) \},
\\
W(\qwr,M) & = \la \qwr, F \ra
\ea
\label{rs-dpe}
\ee
where $0 \leq k \leq M-1$.

\begin{theorem}  \label{thm:rs}
Let $W(\qwr,k)$, $0 \leq k \leq M$, be the solution of the dynamic
programming equation \er{rs-dpe}. (i) Then for any $K
\in \cK$ we have
\be
W(\qwr,k) \leq J^\mu_{\qwr,k}(K)
\label{WJ-1}
\ee
(ii) Assume in addition that the minimizer
\be
\ba{l}
\hat\bu^\ast(\qwr,k)
\\ \in \ds \argmin_{u \in \bU}  \{ \sum_{y \in \bY} W(
\Lambda_{\Gamma, R} (u,y)  \qwr , k+1) p_R(y
\vert u,
\qwr) \}
\ea
\label{rs-u-star}
\ee
exists for all $\qwr$, $0 \leq k \leq M-1$. Then the separation
structure controller $K^{\hat\bu^\ast}_{\qwr_0}$ defined by
\er{rs-u-star} is optimal for problem \er{rs-cost-2}, i.e.
$J^\mu_{\qwr_0,0}(K) \geq
J^\mu_{\qwr_0,0}(K^{\hat\bu^\ast}_{\qwr_0})$ for all $K
\in \cK$.
\end{theorem}

\begin{proof}
We prove part (i) by induction. Let $K \in \cK$. For $k=M$, we
have
$$
W(\qwr,M) = \la \qwr, F \ra = J^\mu_{\qwr,M}(K)
$$
so \er{WJ-1} holds for $k=M$. Next, we assume \er{WJ-1} holds for
$k+1$, i.e.
\be
W(\qwr,k+1) \leq J^\mu_{\qwr,k+1}(K)
\label{WJ-2}
\ee
Now by \er{rs-dpe}, \er{WJ-2} and
\er{trans-prob-L}
$$
\ba{rl}
& W(\qwr,k) \\ & \leq \ds  \sum_{y_{k+1} \in \bY} W(
\Lambda_{\Gamma, R} (u_k,y_{k+1})  \qwr , k+1) p_R(y_{k+1}
\vert u_k,
\qwr)
\\
& \leq  \ds \sum_{y_{k+1} \in \bY} J^\mu_{ \Lambda_{\Gamma, R}
(u_k,y_{k+1})  \qwr  ,k+1}(K)p_R(y_{k+1}
\vert u_k,
\qwr)
\\
& = \ds \sum_{y_{k+1} \in \bY}
\sum_{ y_{k+2,M} \in \bY^{M-(k+1)}}
\la
\Lambda_{\Gamma, R} (u_k,y_{k+1})  \qwr
, G_{k+1} \ra .
\\
& \hspace{5.0cm} .p_R(y_{k+1}
\vert u_k,
\qwr)
\\
& = \ds \sum_{y_{k+1} \in \bY}
\sum_{ y_{k+2,M} \in \bY^{M-(k+1)}}
\la \qwr,
%\\
%& \hspace{4.0cm} .
\Gamma^\dagger_R(u_k,y_{k+1}) G_{k+1} \ra
\\
& = J^\mu_{\qwr,k}(K)
\ea
$$
as required.

Part (ii) follows from the proof of part (i), with $k=0$, since at
every step we have equality and so
$$
W(\qwr_0,0) = J^\mu_{\qwr_0,0}(K^{\hat\bu^\ast}_{\qwr_0})
$$
Hence $J^\mu_{\qwr_0,0}(K) \geq
J^\mu_{\qwr_0,0}(K^{\hat\bu^\ast}_{\qwr_0})$ for all $K
\in \cK$. The real multiplicative homogeneity property of $G_k$
has been used here also.
\qed
\end{proof}

\begin{figure}[htb]
\begin{center}
\setlength{\unitlength}{1800sp}%
\begingroup\makeatletter\ifx\SetFigFont\undefined%
\gdef\SetFigFont#1#2#3#4#5{%
  \reset@font\fontsize{#1}{#2pt}%
  \fontfamily{#3}\fontseries{#4}\fontshape{#5}%
  \selectfont}%
\fi\endgroup%
\begin{picture}(7224,4749)(889,-4798)
\thinlines
{\color[rgb]{0,0,0}\put(3001,-1561){\framebox(3000,1500){}}
}%
{\color[rgb]{0,0,0}\put(2101,-3961){\framebox(2100,1500){}}
}%
{\color[rgb]{0,0,0}\put(4801,-3961){\framebox(2100,1500){}}
}%
{\color[rgb]{0,0,0}\put(4801,-3211){\vector(-1, 0){600}}
}%
{\color[rgb]{0,0,0}\put(2101,-3211){\line(-1, 0){1200}}
\put(901,-3211){\line( 0, 1){2400}}
\put(901,-811){\vector( 1, 0){2100}}
}%
{\color[rgb]{0,0,0}\put(6001,-811){\line( 1, 0){2100}}
\put(8101,-811){\line( 0,-1){2400}}
\put(8101,-3211){\vector(-1, 0){1200}}
}%
{\color[rgb]{0,0,0}\put(1801,-4786){\dashbox{60}(5400,2700){}}
}%
\put(1951,-1111){\makebox(0,0)[lb]{\smash{\SetFigFont{10}{14.4}{\familydefault}{\mddefault}{\updefault}{\color[rgb]{0,0,0}$u$}%
}}}
\put(6826,-1111){\makebox(0,0)[lb]{\smash{\SetFigFont{10}{14.4}{\familydefault}{\mddefault}{\updefault}{\color[rgb]{0,0,0}$y$}%
}}}
\put(1876,-661){\makebox(0,0)[lb]{\smash{\SetFigFont{10}{14.4}{\familydefault}{\mddefault}{\updefault}{\color[rgb]{0,0,0}input}%
}}}
\put(6751,-586){\makebox(0,0)[lb]{\smash{\SetFigFont{10}{14.4}{\familydefault}{\mddefault}{\updefault}{\color[rgb]{0,0,0}output}%
}}}
\put(3426,-1261){\makebox(0,0)[lb]{\smash{\SetFigFont{10}{14.4}{\familydefault}{\mddefault}{\updefault}{\color[rgb]{0,0,0}eqn. \er{system}}%
}}}
\put(3426,-886){\makebox(0,0)[lb]{\smash{\SetFigFont{10}{14.4}{\familydefault}{\mddefault}{\updefault}{\color[rgb]{0,0,0}state $\qw_k$}%
}}}
\put(3426,-436){\makebox(0,0)[lb]{\smash{\SetFigFont{10}{14.4}{\familydefault}{\mddefault}{\updefault}{\color[rgb]{0,0,0}physical system}%
}}}
\put(2701,-2911){\makebox(0,0)[lb]{\smash{\SetFigFont{10}{14.4}{\familydefault}{\mddefault}{\updefault}{\color[rgb]{0,0,0}control}%
}}}
\put(2701,-3511){\makebox(0,0)[lb]{\smash{\SetFigFont{10}{14.4}{\familydefault}{\mddefault}{\updefault}{\color[rgb]{0,0,0}$\hat\bu^\ast(\qwr_k,k)$}%
}}}
\put(5326,-2911){\makebox(0,0)[lb]{\smash{\SetFigFont{10}{14.4}{\familydefault}{\mddefault}{\updefault}{\color[rgb]{0,0,0}filter}%
}}}
\put(5326,-3286){\makebox(0,0)[lb]{\smash{\SetFigFont{10}{14.4}{\familydefault}{\mddefault}{\updefault}{\color[rgb]{0,0,0}state $\qwr_k$}%
}}}
\put(5326,-3661){\makebox(0,0)[lb]{\smash{\SetFigFont{10}{14.4}{\familydefault}{\mddefault}{\updefault}{\color[rgb]{0,0,0}eqn. \er{rs-is-dyn}}%
}}}
\put(3001,-4486){\makebox(0,0)[lb]{\smash{\SetFigFont{10}{14.4}{\familydefault}{\mddefault}{\updefault}{\color[rgb]{0,0,0}feedback controller $K^{\hat\bu^\ast}_{\qwr_0}$}%
}}}
\end{picture}

\caption{Optimal multiplicative/risk-sensitive controller $K^{\hat\bu^\ast}_{\qwr_0}$ showing separation structure and states of the
physical system $\qw_k$ and filter $\qwr_k$.}
\label{fig:fb-rs}
\end{center}
\end{figure}
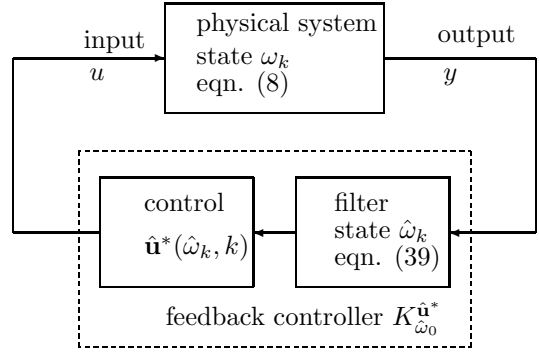

\begin{remark}   \label{rmk:rs-fb}
Note that the optimal multiplicative cost/risk-sensitive
controller $K^{\hat\bu^\ast}_{\qwr_0}$ determined by
\er{rs-u-star} feeds back the unnormalized conditional state $\qwr_k$, given by the
modified SME \er{rs-is-dyn}, Figure
\ref{fig:fb-rs}. This means that from the point of
view of optimal risk-sensitive or multiplicative  control, the
best thing to do involves use of a dynamical quantity that not
only contains {\em knowledge} (as measured by the controller) of
the quantum system, but also contains information about the {\em
purpose} of the controller. This should be contrasted with the
risk-neutral case, section \ref{sec:rn}. Note in particular that
the modified SME
\er{rs-is-dyn} is no longer the optimal filter from the point of
view of seeking the best estimate of observables (c.f. Remark
\ref{rmk:rn-fb}). Further, the concept of dual control (Remark
\ref{rmk:dual}) has greater weight here, since the cost $R(u)$
appears explicitly in the controller dynamics
\er{rs-is-dyn}---while the optimal multiplicative cost/risk-sensitive controller has a
separation structure, in the sense of a decomposition into a
dynamical filter part and static control part, the task of
estimation is not separated from the task of control, Figure
\ref{fig:fb-rs}.
We emphasize that the multiplicative cost/risk-sensitive
conditional state is defined only in the context of these specific
control objectives, where they are used in specific feedback
situations.
\qed
\end{remark}

%%%%%%%%%%%%%%%%%%%%%%%%%%%%%%%%%%%%%%%%%%%%%%%%%%%%%%%%%%%%%%
%%%%%%%%%%%%%%%%%%%%%%%%%%%%%%%%%%%%%%%%%%%%%%%%%%%%%%%%%%%%%%

\begin{example}   \label{eg:2-d-5} (Two-state system with feedback, Example \ref{eg:2-d-4} continued.)
We now consider the risk-sensitive optimal control problem for the
two-state example, with operator valued cost $R(u)$ defined by
\er{r-rs-2}
where $L(u)$ is given by \er{eg1-cf}. If we write the density
matrix as  \er{density-2d}, then
$$
R(u)\qwr =  \frac{e^\mu\qwr_{11} + \qwr_{22}}{\qwr_{11} +
\qwr_{22}} e^{\mu c(u)} \left( \ba{cc} \qwr_{11} & \qwr_{12} \\
\qwr_{12}^\ast & \qwr_{22} \ea
\right) .
$$
The risk-sensitive controlled transfer operators $\Gamma_R(u,y)$
are defined by
$$
\ba{rl}
\Gamma_R(u,y)\qwr & = \ \Gamma(u,y)R(u)\qwr
\\
& = \frac{\la \qwr, e^{\mu L(u)} \ra}{\la \qwr, 1
\ra} \Gamma(u,y)\qwr .
\ea
$$
Explicitly, we have
$$
\ba{rl}
\Gamma_R(0,-1)\qwr & = \frac{e^\mu\qwr_{11} + \qwr_{22}}{\qwr_{11} +
\qwr_{22}}  \left( \ba{cc} (1-\alpha)\qwr_{11} & 0 \\ 0 &  \alpha \qwr_{22} \ea \right)
\\
\Gamma_R(0,1)\qwr & = \frac{e^\mu\qwr_{11} + \qwr_{22}}{\qwr_{11} +
\qwr_{22}} \left( \ba{cc} \alpha \qwr_{11} & 0 \\ 0 &  (1-\alpha) \qwr_{22} \ea \right)
\\
\Gamma_R(1,-1)\qwr & = \frac{e^\mu\qwr_{11} + \qwr_{22}}{\qwr_{11} +
\qwr_{22}} e^{\mu p} \left( \ba{cc} (1-\alpha)\qwr_{22} & 0 \\ 0 &  \alpha \qwr_{11} \ea \right)
\\
\Gamma_R(1,1)\qwr & = \frac{e^\mu\qwr_{11} + \qwr_{22}}{\qwr_{11} +
\qwr_{22}} e^{\mu p} \left( \ba{cc} \alpha\qwr_{22} & 0 \\ 0 &  (1-\alpha) \qwr_{11} \ea
\right).
\ea
$$

The dynamic programming equation \er{rs-dpe} is solved and the
optimal feedback controls are found as follows. First, for $k=M=2$
we have
$$
W(\qwr, 2) = \la \qwr, e^{\mu X^2} \ra = e^\mu \qwr_{11}+\qwr_{22}
$$
and then for $k=1$
$$
W(\qwr,1) = \min[ W_0(\qwr,1), W_0(\qwr,1) ]
$$
where $W_0(\qwr,1)$ and $W_1(\qwr,1)$ are given in Appendix
\ref{app:sg-eg}, and
$$
\hat\bu^\ast(\qw,1) = \left\{  \ba{rl}
0 & \ \text{if} \ W_0(\qw,1) \leq W_1(\qw,1)
\\
1 & \ \text{if} \ W_0(\qw,1) > W_1(\qw,1) .
 \ea \right.
$$
Next, for $k=0$,
$$
W(\qwr,0) = \min[ W_0(\qwr,0), W_0(\qwr,0) ]
$$
where $W_0(\qwr,0)$ and $W_1(\qwr,0)$ are given in Appendix
\ref{app:sg-eg}, and
$$
\hat\bu^\ast(\qw,0) = \left\{  \ba{rl}
0 & \ \text{if} \ W_0(\qw,0) \leq W_1(\qw,0)
\\
1 & \ \text{if} \ W_0(\qw,0) > W_1(\qw,0) .
 \ea \right.
$$

The optimal feedback controller is given by
$$
u_0=K^{\hat\bu^\ast}_{\qwr_0,0} = \hat\bu^\ast(\qwr_0,0), \ \
u_1=K^{\hat\bu^\ast}_{\qwr_0,1}(y_1) =
\hat\bu^\ast(
\qw_1,1)
$$
where $\qwr_1 = \Lambda_{\Gamma,R}(u_0,y_1)\qwr_0$. {\em Again, we
see the separation structure, where here the control $u_1$ depends
on $y_1$ through the unnormalized conditional state $\qwr_1$.} A
physical implementation of this controller is shown in Figure
\ref{fig:sg-2-rs}.

%%%%%%%%%%%%%%%%%%%%%%%%%%%%%%%%%%%%%%%%%%%%%%%%%%%%%%%%%%%%%%
\begin{figure}[htb]
\begin{center}
\setlength{\unitlength}{1600sp}%
\begingroup\makeatletter\ifx\SetFigFont\undefined%
\gdef\SetFigFont#1#2#3#4#5{%
  \reset@font\fontsize{#1}{#2pt}%
  \fontfamily{#3}\fontseries{#4}\fontshape{#5}%
  \selectfont}%
\fi\endgroup%
\begin{picture}(9324,8424)(2089,-6973)         %\begin{picture}(9324,8424)(2089,-6973)
{\color[rgb]{0,0,0}\thinlines
\put(5251,-2611){\circle{150}}
}%
{\color[rgb]{0,0,0}\put(5251,-2011){\circle{150}}
}%
{\color[rgb]{0,0,0}\put(8401,-2761){\framebox(900,900){}}
}%
{\color[rgb]{0,0,0}\put(3751,-2761){\framebox(900,900){}}
}%
{\color[rgb]{0,0,0}}%
{\color[rgb]{0,0,0}}%
{\color[rgb]{0,0,0}\put(7801,-2311){\vector( 1, 0){600}}
}%
{\color[rgb]{0,0,0}\put(2401,-2311){\vector( 1, 0){1350}}
}%
{\color[rgb]{0,0,0}}%
{\color[rgb]{0,0,0}\put(9301,-2611){\line( 1, 0){1350}}
\put(10651,-2611){\vector( 0, 1){300}}
}%
{\color[rgb]{0,0,0}\put(9301,-2011){\line( 1, 0){1350}}
\put(10651,-2011){\vector( 0,-1){300}}
}%
{\color[rgb]{0,0,0}\put(10651,-2311){\vector( 1, 0){450}}
}%
{\color[rgb]{0,0,0}}%
{\color[rgb]{0,0,0}\put(6526,-2236){\framebox(750,525){}}
}%
{\color[rgb]{0,0,0}\put(6526,-2986){\framebox(750,525){}}
}%
{\color[rgb]{0,0,0}\put(4651,-2011){\vector( 1, 0){1875}}
}%
{\color[rgb]{0,0,0}\put(4651,-2611){\vector( 1, 0){1875}}
}%
{\color[rgb]{0,0,0}\put(7276,-2611){\line( 1, 0){525}}
\put(7801,-2611){\vector( 0, 1){300}}
}%
{\color[rgb]{0,0,0}\put(7276,-2011){\line( 1, 0){525}}
\put(7801,-2011){\vector( 0,-1){300}}
}%
{\color[rgb]{0,0,0}\put(6001,-3286){\dashbox{60}(4050,1800){}}
}%
{\color[rgb]{0,0,0}\put(3001,-3286){\dashbox{60}(2400,1800){}}
}%
{\color[rgb]{0,0,0}\put(2201,-4411){\dashbox{60}(9100,3300){}} %60,9300,3300
}%
{\color[rgb]{0,0,0}\put(5401,-5911){\vector( 1, 0){750}}
}%
{\color[rgb]{0,0,0}\put(5251,-2686){\vector( 0,-1){2775}}
}%
{\color[rgb]{0,0,0}\put(6901,-5461){\vector( 0, 1){2475}}
}%
{\color[rgb]{0,0,0}\put(5401,389){\vector( 1, 0){750}}
}%
{\color[rgb]{0,0,0}\put(5251,-1936){\vector( 0, 1){1875}}
}%
{\color[rgb]{0,0,0}\put(6901,-61){\vector( 0,-1){1650}}
}%
{\color[rgb]{0,0,0}\put(2851,-61){\framebox(2550,900){}}
}%
{\color[rgb]{0,0,0}\put(6151,-61){\framebox(1800,900){}}
}%
{\color[rgb]{0,0,0}\put(2551,-361){\dashbox{60}(5700,1800){}}
}%
{\color[rgb]{0,0,0}\put(2851,-6361){\framebox(2550,900){}}
}%
{\color[rgb]{0,0,0}\put(6076,-6361){\framebox(1800,900){}}
}%
{\color[rgb]{0,0,0}\put(2551,-6961){\dashbox{60}(5700,1800){}}
}%
\put(8476,-2386){\makebox(0,0)[lb]{\smash{\SetFigFont{8}{14.4}{\familydefault}{\mddefault}{\updefault}{\color[rgb]{0,0,0}M-$\alpha$}%
}}}
\put(3826,-2386){\makebox(0,0)[lb]{\smash{\SetFigFont{8}{14.4}{\familydefault}{\mddefault}{\updefault}{\color[rgb]{0,0,0}M-$\alpha$}%
}}}
\put(4876,-2436){\makebox(0,0)[lb]{\smash{\SetFigFont{10}{14.4}{\familydefault}{\mddefault}{\updefault}{\color[rgb]{0,0,0}$1$}%
}}}
\put(9526,-2436){\makebox(0,0)[lb]{\smash{\SetFigFont{10}{14.4}{\familydefault}{\mddefault}{\updefault}{\color[rgb]{0,0,0}$1$}%
}}}
\put(4676,-3036){\makebox(0,0)[lb]{\smash{\SetFigFont{10}{14.4}{\familydefault}{\mddefault}{\updefault}{\color[rgb]{0,0,0}$-1$}%
}}}
\put(9526,-3036){\makebox(0,0)[lb]{\smash{\SetFigFont{10}{14.4}{\familydefault}{\mddefault}{\updefault}{\color[rgb]{0,0,0}$-1$}%
}}}
\put(2401,-3961){\makebox(0,0)[lb]{\smash{\SetFigFont{10}{14.4}{\familydefault}{\mddefault}{\updefault}{\color[rgb]{0,0,0}$\qw_0$}%
}}}
\put(10576,-3961){\makebox(0,0)[lb]{\smash{\SetFigFont{10}{14.4}{\familydefault}{\mddefault}{\updefault}{\color[rgb]{0,0,0}$\rho_2$}%
}}}
\put(9901,-3961){\makebox(0,0)[lb]{\smash{\SetFigFont{10}{14.4}{\familydefault}{\mddefault}{\updefault}{\color[rgb]{0,0,0}$\tilde\qw_2$}%
}}}
\put(5476,-3961){\makebox(0,0)[lb]{\smash{\SetFigFont{10}{14.4}{\familydefault}{\mddefault}{\updefault}{\color[rgb]{0,0,0}$\qw_1$}%
}}}
\put(6601,-2111){\makebox(0,0)[lb]{\smash{\SetFigFont{10}{14.4}{\familydefault}{\mddefault}{\updefault}{\color[rgb]{0,0,0}$T^{u_1}$}%
}}}
\put(6601,-2861){\makebox(0,0)[lb]{\smash{\SetFigFont{10}{14.4}{\familydefault}{\mddefault}{\updefault}{\color[rgb]{0,0,0}$T^{u_1}$}%
}}}
\put(8226,-4336){\makebox(0,0)[lb]{\smash{\SetFigFont{10}{14.4}{\familydefault}{\mddefault}{\updefault}{\color[rgb]{0,0,0}physical system}%
}}}
\put(6226,-5986){\makebox(0,0)[lb]{\smash{\SetFigFont{10}{14.4}{\familydefault}{\mddefault}{\updefault}{\color[rgb]{0,0,0}$\hat\bu^\ast(\qwr_1,1)$}%
}}}
\put(3001,-5986){\makebox(0,0)[lb]{\smash{\SetFigFont{10}{14.4}{\familydefault}{\mddefault}{\updefault}{\color[rgb]{0,0,0}$\qw_0 \to\qwr_1^{(0,-1)}$}%
}}}
\put(6301,314){\makebox(0,0)[lb]{\smash{\SetFigFont{10}{14.4}{\familydefault}{\mddefault}{\updefault}{\color[rgb]{0,0,0}$\hat\bu^\ast(\qwr_1,1)$}%
}}}
\put(3126,314){\makebox(0,0)[lb]{\smash{\SetFigFont{10}{14.4}{\familydefault}{\mddefault}{\updefault}{\color[rgb]{0,0,0}$\qw_0\to\qwr_1^{(0,1)}$}%
}}}
\put(3751,1064){\makebox(0,0)[lb]{\smash{\SetFigFont{10}{14.4}{\familydefault}{\mddefault}{\updefault}{\color[rgb]{0,0,0}filter}%
}}}
\put(6426,1064){\makebox(0,0)[lb]{\smash{\SetFigFont{10}{14.4}{\familydefault}{\mddefault}{\updefault}{\color[rgb]{0,0,0}control}%
}}}
\put(4001,-736){\makebox(0,0)[lb]{\smash{\SetFigFont{10}{14.4}{\familydefault}{\mddefault}{\updefault}{\color[rgb]{0,0,0}$y_1=1$}%
}}}
\put(6301,-736){\makebox(0,0)[lb]{\smash{\SetFigFont{10}{14.4}{\familydefault}{\mddefault}{\updefault}{\color[rgb]{0,0,0}$u_1$}%
}}}
\put(6301,-4861){\makebox(0,0)[lb]{\smash{\SetFigFont{10}{14.4}{\familydefault}{\mddefault}{\updefault}{\color[rgb]{0,0,0}$u_1$}%
}}}
\put(3726,-4861){\makebox(0,0)[lb]{\smash{\SetFigFont{10}{14.4}{\familydefault}{\mddefault}{\updefault}{\color[rgb]{0,0,0}$y_1=-1$}%
}}}
\put(3751,-6736){\makebox(0,0)[lb]{\smash{\SetFigFont{10}{14.4}{\familydefault}{\mddefault}{\updefault}{\color[rgb]{0,0,0}filter}%
}}}
\put(6426,-6736){\makebox(0,0)[lb]{\smash{\SetFigFont{10}{14.4}{\familydefault}{\mddefault}{\updefault}{\color[rgb]{0,0,0}control}%
}}}
\end{picture}

\caption{Physical realization of the two stages of the two-state system with feedback using the optimal
risk-sensitive controller $K^{\bar\bu^\ast}_{\qwr_0}$ (with
$\qwr_0=\qw_0$ given by
\er{eg1-w0}, we have  $u_0=\hat\bu^\ast(\qw_0,0)=0$, $u_1=\hat\bu^\ast(\qwr_1,1)$).}
\label{fig:sg-2-rs}
\end{center}
\end{figure}
%%%%%%%%%%%%%%%%%%%%%%%%%%%%%%%%%%%%%%%%%%%%%%%%%%%%%%%%%%%%%%

To see the effect of the risk-sensitive controller, consider the
initial state $\qwr_0=\qw_0$ given by \er{eg1-w0}, and parameter
values $\alpha=0.25$, $\mu=2$. We find that
$W_0(\qwr_0,0)=W_1(\qwr_0,0)$, and hence we take
$u_0=\hat\bu^\ast(\qwr_0,0)=0$.

If $y_1=-1$ is measured, we find that
$$
\ba{rl}
\qwr_1 & = \left( \ba{cc}
3.1459 & 0 \\ 0 & 1.04863 \ea\right),
\\
\qw_1 & = \left( \ba{cc}
0.75 & 0 \\ 0 & 0.25 \ea\right) \ \text{with prob.} \ 0.5,
\ea
$$
and
\be
u_1 = \hat\bu^\ast(\qwr_1,1) = \left\{  \ba{rl} 1 & \ \text{if} \
0
\leq p
\leq 0.4
\\
 0 & \ \text{if}\  p > 0.4
\ea \right.
\label{eg1-u-star-p-rs}
\ee
If, on the other hand, $y_1=1$ is measured, we find that
$$
\ba{rl}
\qwr_1 & = \left( \ba{cc}
1.04863 & 0 \\ 0 & 3.1459 \ea\right),
\\
\qw_1 & = \left( \ba{cc}
0.25 & 0 \\ 0 & 0.75 \ea\right) \ \text{with prob.} \ 0.5,
\ea
$$
and
$$
u_1 = \hat\bu^\ast(\qwr_1,1) = 0
$$
for any value of $p$. When the control cost $p=0.2$, the final
(non-selective) state is given by
$$
\rho_2=  \left( \ba{cc}
0.25 & 0 \\ 0 & 0.75 \ea\right) = 0.25 \vert -1 \ra\la -1 \vert
+0.75 \vert 1 \ra\la 1 \vert
$$
This state does not equal the desired pure state $\vert 1 \ra\la 1
\vert$, a reflection of the level of measurement uncertainty
$\alpha=0.25$ and the presence of a non-zero control penalty.

To compare with the risk-neutral version of this problem, we find
that the threshold in \er{eg1-u-star-p-rs} for the risk-neutral
problem is $p=0.75$. This means that for a larger range of values
of the control cost $p$, the risk-neutral controller will be
active, i.e. select $u=1$ than is the case for the risk-sensitive
controller. This is consistent with the description of the example
in \cite{FGM94,CM95}  where the risk-neutral controller is more
aggressive than the risk-sensitive controller.
\qed
\end{example}

We conclude with an example which indicates the likely robustness
properties of the risk-sensitive controller and the relationship
between the risk-neutral and risk-sensitive problems.

\begin{example} \label{eg:simple}
We consider the risk-sensitive cost functional \er{r-rs-1}, where
where the operator valued cost $R(u)$ has the form \er{r-rs-2},
and $F=e^{\mu N}$.

{\bf Robustness.} To describe the robustness properties of the
risk-sensitive controller, we follow \cite{DJP00} and make use of
the following general convex duality formula (see, e.g.
\cite[Chapter 1.4]{DE97}):
\be
\log \E_{\bP} [ e^f ] = \sup_{\bQ} \{ \E_{\bQ} [f] - RE(\bQ \pa
\bP) \}
\label{dua}
\ee
where $\bP$ and $\bQ$ are probability distributions\footnote{We
require that $\bQ$ is absolutely continuous with respect to
$\bP$.}, and where the relative entropy is defined by (see, e.g.,
\cite[Chapter 11]{NC00})
$$
RE(\bQ \pa \bP) = \E_{\bQ} [ \log \frac{d
\bQ}{d \bP} ] .
$$
To apply formula \er{dua}, we proceed as follows. Let
$\Gamma_{nom}$ be the nominal operator  used for design of the
optimal risk-sensitive controller, here denoted $\hat
K^\ast_{nom}$. Together, $\Gamma_{nom}$ and $\hat K^\ast_{nom}$
determine a probability distribution, here denoted $\bP_{nom}$. In
reality, the nominal $\Gamma_{nom}$ need not equal the operator
for the \lq\lq{true}\rq\rq \ system, denoted $\Gamma_{true}$. The
controller $\hat K^\ast_{nom}$ is applied to the true system,
resulting in a probability distribution $\bP_{true}$\footnote{We
assume that the true distribution $\bP_{true}$ is absolutely
continuous with respect to the nominal distribution $\bP_{nom}$.}.

We write $\mu=1/\gamma^2$, and apply \er{dua} to obtain the
following inequality ($\bP=\bP_{nom}$, $\bQ=\bP_{true}$):
$$
\ba{l}
\gamma^2 \log \E_{\bP_{true}} [ \prod_{k=0}^{M-1} \la
\qw_k,e^{\mu L(u_k)}\ra
\la \qw_M, e^{\mu N} \ra  ]
\\
\geq  \E_{\bP_{true}} [\gamma^2 \log (\prod_{k=0}^{M-1} \la
\qw_k,e^{\mu L(u_k)}\ra
\la \qw_M, e^{\mu N} \ra  ) ]
\\
\ \ \  \  - \gamma^2 RE(\bP_{true} \pa
\bP_{nom})
\\
= \E_{\bP_{true}} [\gamma^2 (\sum_{k=0}^{M-1} \log \la
\qw_k,e^{\mu L(u_k)}\ra
\\ \hspace{2.0cm} +\log
\la \qw_M, e^{\mu N} \ra  ) ]
\\
\ \hspace{2.5cm}  - \gamma^2 RE(\bP_{true} \pa
\bP_{nom})
\\
\geq  \E_{\bP_{true}} [\sum_{k=0}^{M-1} \la
\qw_k,L(u_k) \ra +
\la \qw_M, N \ra  ) ]
\\
\ \ \  \  - \gamma^2 RE(\bP_{true} \pa
\bP_{nom})
\ea
$$
This implies the important bound:
\be
\ba{l}
J^{rn}_{\bP_{true}}(\bar K^\ast_{nom})
\\
\ \ \leq \gamma^2 \log J^{rs, \gamma^2}_{\bP_{nom}}(\hat K^\ast_{nom})
+ \gamma^2 RE(\bP_{true} \pa
\bP_{nom})
\ea
\label{robust}
\ee
The LHS of \er{robust} is the risk-neutral cost criterion
\er{rn-value-def}, evaluated using the true system model $\bP_{true}$ and the
controller $\hat K^\ast_{nom}$ designed using the nominal model
$\bP_{nom}$. Inequality
\er{robust} bounds this cost by two terms, the first term is
related to the optimal risk-sensitive cost \er{r-rs-1}, while the
second is the relative entropy term, which is a measure of the
\lq\lq{distance}\rq\rq \ between the true and nominal systems.
The number $\gamma^2 = 1/\mu > 0$ is a \lq\lq{robustness
gain}\rq\rq
\ parameter, which we would like to be as small as possible for maximum robustness,
 as in $H^\infty$ robust control, \cite{GL95}, where the relative
 entropy term is a measure of the \lq\lq{energy}\rq\rq \ in the
 disturbance or uncertainty.
This shows that the risk-sensitive controller enjoys
good performance, as measured by the risk-neutral criterion, under
nominal conditions $(\bP_{true} =
\bP_{nom})$, and acceptable performance in other than
nominal conditions $(\bP_{true} \neq
\bP_{nom})$, as implied by the bound. In summary,
risk-sensitive controllers enjoy enhanced robustness (recall
Remark \ref{rmk:robust-1}).

{\bf Relationship between the  risk-neutral and risk-sensitive
value functions.} We indicate briefly how the results of
\cite[Theorem 5.5]{JBE94} apply in the present context.
Indeed, the reader may check that for small $\mu >0$ one has
$$
\frac{1}{\mu} \log \frac{\la \qwr, \exp(\mu N) \ra}{\la \qwr,1\ra}
\approx \frac{\la \qwr, N\ra}{\la \qwr, 1\ra}
$$
This suggests the relation
\be
\lim_{\mu \downarrow 0} \frac{1}{\mu} \log \frac{W(\qwr,k)}{\la \qwr,1\ra} =
\frac{V(\qwr,k)}{\la \qwr,1\ra},
\label{rs-rn-limit}
\ee
which says that a logarithmic  risk-sensitive optimal cost tends
to the optimal risk-neutral cost as the parameter $\mu \to 0$,  as
might be expected.
\qed
\end{example}

%%%%%%%%%%%%%%%%%%%%%%%%%

\appendix

\section{Formulas for the Two-State System with Feedback Example}
\label{app:sg-eg}

The following quantities were used in the solution of the
risk-neutral problem, Example \ref{eg:2-d-4}:
$$
V_0(\qw,1) = \qw_{11}, \ \ V_1(\qw,1) =\qw_{22}+p
$$
$$
\ba{l}
V_0(\qw,0) = \qw_{11} + \min[\alpha \qw_{11}, p +
\qw_{22} - \alpha \qw_{22}]
\\ \hspace{2.5cm} + \min[\qw_{11} - \alpha \qw_{11}, p + \alpha
\qw_{22}]
\ea
$$
$$
\ba{l}
V_1(\qw,0) = p + \alpha \qw_{11} + \qw_{22} - \alpha \qw_{22}
    \\ \hspace{2.5cm} +
\min[\alpha \qw_{11}, p + \qw_{22} - \alpha \qw_{22}]
\\ \hspace{2.5cm} +
      \min[p + \alpha \qw_{11}, \qw_{22} - \alpha \qw_{22}]
\ea
$$

The following quantities were used in the solution of the
risk-sensitive problem, Example \ref{eg:2-d-5}:
\begin{widetext}
$$
W_0(\qwr,1) = \frac{{( e^{\mu}\,\qwr_{11} + \qwr_{22} ) }^2}
  {\qwr_{11} + \qwr_{22}}, \ \
W_1(\qwr,1) = \frac{e^{\mu\,p}\,( \qwr_{11}\,\qwr_{22} +
      e^{2\,\mu}\,\qwr_{11}\,\qwr_{22} +
      e^{\mu}\,( {\qwr_{11}}^2 + {\qwr_{22}}^2 )
      ) }{\qwr_{11} + \qwr_{22}}
$$
$$
\ba{rl}
W_0(\qwr,0) & =
\min [-( \frac{( e^{\mu}\,\qwr_{11} +
          \qwr_{22} ) \,
        {( ( -1 + \alpha ) \,e^{\mu}\,\qwr_{11}
        -    \alpha\,\qwr_{22} ) }^2}{( \qwr_{11} +
          \qwr_{22} ) \,
        ( ( -1 + \alpha ) \,\qwr_{11} -
          \alpha\,\qwr_{22} ) } ) ,
          \\
  & \ \ \ -( \frac{e^{\mu\,p}\,( e^{\mu}\,\qwr_{11} +
          \qwr_{22} ) \,
        ( -( ( -1 + \alpha ) \,\alpha\,\qwr_{11}\,
             \qwr_{22} )  -
          ( -1 + \alpha ) \,\alpha\,e^{2\,\mu}\,\qwr_{11}\,
           \qwr_{22} +
          e^{\mu}\,( {( -1 + \alpha ) }^2\,
              {\qwr_{11}}^2 + \alpha^2\,{\qwr_{22}}^2 )
          ) }{( \qwr_{11} + \qwr_{22} ) \,
        ( ( -1 + \alpha ) \,\qwr_{11} -
          \alpha\,\qwr_{22} ) } ) ]
          \\
      & \ \ \   +
  \min ([\frac{( e^{\mu}\,\qwr_{11} + \qwr_{22} )
        \,{( \alpha\,e^{\mu}\,\qwr_{11} + \qwr_{22} -
          \alpha\,\qwr_{22} ) }^2}{( \qwr_{11} +
        \qwr_{22} ) \,
      ( \alpha\,( \qwr_{11} - \qwr_{22} )  +
        \qwr_{22} ) },
        \\
  & \ \ \ \ \ \frac{e^{\mu\,p}\,( e^{\mu}\,\qwr_{11} +
        \qwr_{22} ) \,
      ( e^{\mu}\,{\qwr_{22}}^2 +
        \alpha\,\qwr_{22}\,( \qwr_{11} +
           e^{2\,\mu}\,\qwr_{11} - 2\,e^{\mu}\,\qwr_{22}
           )  + \alpha^2\,( -( \qwr_{11}\,
              \qwr_{22} )  -
           e^{2\,\mu}\,\qwr_{11}\,\qwr_{22} +
           e^{\mu}\,( {\qwr_{11}}^2 + {\qwr_{22}}^2
              )  )  ) }{( \qwr_{11} +
        \qwr_{22} ) \,
      ( \alpha\,( \qwr_{11} - \qwr_{22} )  +
        \qwr_{22} ) }]
\ea
$$
$$
\ba{rl}
W_1(\qwr,0) & =
\min [-( \frac{e^{\mu\,p}\,
        ( e^{\mu}\,\qwr_{11} + \qwr_{22} ) \,
        {( \qwr_{11} - \alpha\,\qwr_{11} +
            \alpha\,e^{\mu}\,\qwr_{22} ) }^2}{( {wr1
           1} + \qwr_{22} ) \,
        ( ( -1 + \alpha ) \,\qwr_{11} -
          \alpha\,\qwr_{22} ) } ) ,
   \\   & \ \ \
   -( \frac{e^{2\,\mu\,p}\,
        ( e^{\mu}\,\qwr_{11} + \qwr_{22} ) \,
        ( -( ( -1 + \alpha ) \,\alpha\,\qwr_{11}\,
             \qwr_{22} )  -
          ( -1 + \alpha ) \,\alpha\,e^{2\,\mu}\,\qwr_{11}\,
           \qwr_{22} +
          e^{\mu}\,( {( -1 + \alpha ) }^2\,
              {\qwr_{11}}^2 + \alpha^2\,{\qwr_{22}}^2 )
          ) }{( \qwr_{11} + \qwr_{22} ) \,
        ( ( -1 + \alpha ) \,\qwr_{11} -
          \alpha\,\qwr_{22} ) } ) ]
          \\
      & \ \ \     +
  \min [\frac{e^{\mu\,p}\,( e^{\mu}\,\qwr_{11} +
        \qwr_{22} ) \,
      {( e^{\mu}\,\qwr_{22} +
          \alpha\,( \qwr_{11} - e^{\mu}\,\qwr_{22} )
          ) }^2}{( \qwr_{11} + \qwr_{22} )
        \,( \alpha\,( \qwr_{11} - \qwr_{22} )  +
        \qwr_{22} ) },
        \\
  & \ \ \ \ \ \  \frac{e^{2\,\mu\,p}\,( e^{\mu}\,\qwr_{11} +
        \qwr_{22} ) \,
      ( e^{\mu}\,{\qwr_{22}}^2 +
        \alpha\,\qwr_{22}\,( \qwr_{11} +
           e^{2\,\mu}\,\qwr_{11} - 2\,e^{\mu}\,\qwr_{22}
           )  + \alpha^2\,( -( \qwr_{11}\,
              \qwr_{22} )  -
           e^{2\,\mu}\,\qwr_{11}\,\qwr_{22} +
           e^{\mu}\,( {\qwr_{11}}^2 + {\qwr_{22}}^2
              )  )  ) }{( \qwr_{11} +
        \qwr_{22} ) \,
      ( \alpha\,( \qwr_{11} - \qwr_{22} )  +
        \qwr_{22} )
         }]
\ea
$$
\end{widetext}

%%%%%%%%%%%%%%%%%%%%%%%%%%%%%%%%%%%%%%%

\bibliographystyle{plain}

\bibliography{mjbib1,qc1}

\begin{thebibliography}{10}

\bibitem{KA70}
K.J. Astrom.
\newblock {\em Introduction to Stochastic Control Theory}.
\newblock Academic Press, New York, 1970.

\bibitem{BJ97}
J.S. Baras and M.R. James.
\newblock Robust and risk-sensitive for finite state machines and hidden markov
  models.
\newblock {\em J. Math. Systems, Estimation and Control}, 7:371--374, 1997.

\bibitem{BB91}
A.~Barchielli and V.P. Belavkin.
\newblock Measurements continuous in time and {\em a posteriori} states in
  quantum mechanics.
\newblock {\em J. Phys. A: Math. Gen.}, 24:1495--1514, 1991.

\bibitem{VPB83}
V.P. Belavkin.
\newblock On the theory of controlling observable quantum systems.
\newblock {\em Automation and Remote Control}, 44(2):178--188, 1983.

\bibitem{VPB01}
V.P. Belavkin.
\newblock Quantum noise, bits and jumps: Uncertainties, decoherence,
  trajectories and filtering.
\newblock {\em Progress in Quantum Electronics}, 25(1):3--50, 2001.

\bibitem{BV85}
A.~Bensoussan and J.H. van Schuppen.
\newblock Optimal control of partially observable stochastic systems with an
  exponential-of-integral performance index.
\newblock {\em SIAM Journal on Control and Optimization}, 23:599--613, 1985.

\bibitem{DB95}
D.P. Bertsekas.
\newblock {\em Dynamic Programming and Optimal Control}, volume~1.
\newblock Athena Scientific, Boston, 1995.

\bibitem{CM95}
S.P. Coraluppi and S.I. Marcus.
\newblock Risk-sensitive and minimax control of discrete-time, finite-state
  markov decision processes.
\newblock {\em Automatica}, 35, 1999.

\bibitem{EBD76}
E.B. Davies.
\newblock {\em Quantum Theory of Open Systems}.
\newblock Academic Press, New York, 1976.

\bibitem{DHJMT00}
A.C. Doherty, S.~Habib, K.~Jacobs, H.~Mabuchi, and S.M Tan.
\newblock Quantum feedback control and classical control theory.
\newblock {\em Phys. Rev. A}, 62:012105, 2000.
\newblock quant-ph/9912107.

\bibitem{DJ99}
A.C. Doherty and K.~Jacobs.
\newblock Feedback-control of quantum systems using continuous
  state-estimation.
\newblock {\em Phys. Rev. A}, 60:2700, 1999.
\newblock quant-ph/9812004.

\bibitem{DJJ00}
A.C. Doherty, K.~Jacobs, and G.~Jungman.
\newblock Information, disturbance and hamiltonian quantum feedback control.
\newblock {\em quant-ph/0006013}, 2000.

\bibitem{DGKF89}
J.C. Doyle, K.~Glover, P.P. Khargonekar, and B.~Francis.
\newblock State-space solutions to the standard ${H}_2$ and ${H}_\infty$
  control problems.
\newblock {\em IEEE Transactions on Automatic Control}, 34(8):831--847, 1989.

\bibitem{DE97}
P.~Dupuis and R.S. Ellis.
\newblock {\em A Weak Convergence Approach to the Theory of Large Deviations}.
\newblock Wiley, New York, 1997.

\bibitem{DJP00}
P.~Dupuis, M.R. James, and I.R. Petersen.
\newblock Robust properties of risk-sensitive control.
\newblock {\em Math. Control, Systems and Signals}, 13:318--332, 2000.

\bibitem{FGM94}
E.~Fernandez-Gaucherand and S.I. Marcus.
\newblock Risk-sensitive optimal control of hidden markov models: Structural
  results.
\newblock {\em IEEE Trans. Automatic Control}, 42(10):1418 --1422, 1997.

\bibitem{GZ00}
C.W. Gardiner and P.~Zoller.
\newblock {\em Quantum Noise}.
\newblock Springer, Berlin, 2000.

\bibitem{GD88}
K.~Glover and J.C. Doyle.
\newblock State-space formulae for all stabilizing controllers that satisfy an
  ${H}_\infty$ norm bound and relations to risk-sensitivity.
\newblock {\em Systems and Control Letters}, pages 167--172, 1988.

\bibitem{GL95}
M.~Green and D.J.N. Limebeer.
\newblock {\em Linear Robust Cotrol}.
\newblock Prentice-Hall, Englewood Cliffs, NJ, 1995.

\bibitem{J73}
D.H. Jacobson.
\newblock Optimal stochastic linear systems with exponential performance
  criteria and their relation to deterministic differential games.
\newblock {\em IEEE Transactions on Automatic Control}, 18(2):124--131, 1973.

\bibitem{JBE94}
M.R. James, J.S. Baras, and R.J. Elliott.
\newblock Risk-sensitive control and dynamic games for partially observed
  discrete-time nonlinear systems.
\newblock {\em IEEE Transactions on Automatic Control}, 39:780--792, 1994.

\bibitem{KV86}
P.R. Kumar and P.~Varaiya.
\newblock {\em Stochastic Systems: Estimation, Identification and Adaptive
  Control}.
\newblock Prentice-Hall, Englewood Cliffs, NJ, 1986.

\bibitem{NC00}
M.A. Nielsen and I.L. Chuang.
\newblock {\em Quantum Computation and Quantum Information}.
\newblock Cambridge University Press, Cambridge, 2000.

\bibitem{VDM01}
F.~Verstraete, A.C. Doherty, and H.~Mabuchi.
\newblock Sensitivity optimization in quantum parameter estimation.
\newblock {\em Phys. Rev. A}, 64(3):032111, 2001.
\newblock quant-ph/0104116.

\bibitem{W81}
P.~Whittle.
\newblock Risk-sensitive linear/ quadratic/ {G}aussian control.
\newblock {\em Advances in Applied Probability}, 13:764--777, 1981.

\bibitem{HW94}
H.~Wiseman.
\newblock {\em Quantum Trajectories and Feedback}.
\newblock PhD thesis, University of Queensland, Brisbane, Australia, 1994.

\bibitem{WM94}
H.~Wiseman and G.J. Milburn.
\newblock Squeezing via feedback.
\newblock {\em Phys. Rev. A}, 49(2):1350--1366, 1994.

\end{thebibliography}

\end{document}